\documentclass[11pt]{article}

\usepackage[preprint]{acl}

\usepackage{times}
\usepackage{latexsym}

\usepackage[T1]{fontenc}

\usepackage{xcolor}
\usepackage{listings}

\usepackage{xcolor}
\usepackage{listings}

\definecolor{pykeyword}{RGB}{0,0,255}
\definecolor{pycomment}{RGB}{0,128,0}
\definecolor{pystring}{RGB}{163,21,21}
\definecolor{linegray}{RGB}{200,200,200}

\lstdefinestyle{base_style}{
    basicstyle=\ttfamily\footnotesize,
    breaklines=true,
    breakindent=0pt,             
    breakatwhitespace=false,
    xleftmargin=0pt,             
    xrightmargin=0pt,
    frame=lines,                 
    framerule=0.5pt,
    rulecolor=\color{linegray},
    keepspaces=true,
    showstringspaces=false,
    aboveskip=10pt,
    belowskip=10pt,
    columns=flexible
}

\lstnewenvironment{pycode}[1][] 
{
    \lstset{
        style=base_style,
        language=Python,
        keywordstyle=\color{pykeyword}\bfseries,
        commentstyle=\color{pycomment}\itshape,
        stringstyle=\color{pystring},
        morekeywords={as,yield,self,True,False,None}, 
        #1 
    }
}{}

\usepackage[utf8]{inputenc}
\usepackage{subcaption}
\usepackage{microtype}

\usepackage{inconsolata}

\usepackage{graphicx}
\usepackage{hyperref}
\usepackage{url}
\usepackage{enumitem}
\usepackage{rotating}
\usepackage{tabularx}
\usepackage{multirow}
\usepackage{amsmath}
\usepackage{listing}
\usepackage{tcolorbox}
\usepackage{wrapfig}
\usepackage{makecell}
\usepackage{tablefootnote}
\usepackage{mdframed}
\usepackage[utf8]{inputenc} 
\usepackage{hyperref}       
\usepackage{url}            
\usepackage{booktabs}       
\usepackage{amsfonts}       
\usepackage{nicefrac}       
\usepackage{microtype}      
\usepackage{xcolor}         
\usepackage{pifont}
\usepackage{graphicx}
\usepackage{multirow}
\usepackage{afterpage}
\usepackage{xcolor}
\usepackage{listings}
\usepackage{csquotes}
\usepackage{multirow}
\usepackage{marvosym}

\title{\textsc{CoreCodeBench}: Decoupling Code Intelligence via Fine-Grained Repository-Level Tasks}

\author{
  \textbf{Lingyue Fu\textsuperscript{1}$^{*,\diamond}$},
  \textbf{Hao Guan\textsuperscript{1}$^{*,\diamond}$},
  \textbf{Bolun Zhang\textsuperscript{1}$^{\diamond}$},
  \textbf{Haowei Yuan\textsuperscript{1}$^{\diamond}$},
  \textbf{Yaoming Zhu\textsuperscript{2}},
  \\
  \textbf{Jun Xu\textsuperscript{2}},
  \textbf{Zongyu Wang\textsuperscript{2}},
  \textbf{Lin Qiu\textsuperscript{2}},
  \textbf{Xunliang Cai\textsuperscript{2}},
  \textbf{Xuezhi Cao\textsuperscript{2}},
  \\
  \textbf{Weiwen Liu\textsuperscript{1}\Letter},
  \textbf{Weinan Zhang\textsuperscript{1}},
  \textbf{Yong Yu\textsuperscript{1}}
\\
  \textsuperscript{1}Shanghai Jiao Tong University,
  \textsuperscript{2}Meituan
  \\
  \small{
  $^*$ Equal contribution \quad
    $^{\diamond}$ Work done while interning at Meituan
  }
  \\
  \small{
    {fulingyue@sjtu.edu.cn, wwliu@sjtu.edu.cn}
  }
}

\begin{document}
\maketitle
\begin{abstract}
The evaluation of Large Language Models (LLMs) for software engineering has shifted towards complex, repository-level tasks. However, existing benchmarks predominantly rely on coarse-grained pass rates that treat programming proficiency as a monolithic capability, obscuring specific cognitive bottlenecks. Furthermore, the static nature of these benchmarks renders them vulnerable to data contamination and performance saturation. To address these limitations, we introduce CoreCodeBench, a configurable repository-level benchmark designed to dissect coding capabilities through atomized tasks. Leveraging our automated framework, CorePipe, we extract and transform Python repositories into a comprehensive suite of tasks that isolate distinct cognitive demands within identical code contexts. Unlike static evaluations, CoreCodeBench supports controllable difficulty scaling to prevent saturation and ensures superior data quality. It achieves a 78.55\% validity yield, significantly surpassing the 31.7\% retention rate of SWE-bench-Verified. Extensive experiments with state-of-the-art LLMs reveal a significant capability misalignment, evidenced by distinct ranking shifts across cognitive dimensions. This indicates that coding proficiency is non-monolithic, as strength in one aspect does not necessarily translate to others. These findings underscore the necessity of our fine-grained taxonomy in diagnosing model deficiencies and offer a sustainable, rigorous framework for evolving code intelligence.
Code of CorePipe framework\footnote{https://github.com/AGI-Eval-Official/CoreCodeBench} and data of CoreCodeBench are available\footnote{https://huggingface.co/collections/tubehhh/corecodebench}.
\end{abstract}

\section{Introduction}
The evaluation of Large Language Models (LLMs) for code has evolved from function-level snippets~\citep{chen2021evaluatinglargelanguagemodels,austin2021programsynthesislargelanguage} to complex, repository-level software engineering tasks, including code generation~\citep{yang2024execrepobenchmultilevelexecutablecode}, program repair~\citep{openai2024swebench} and unit test generation~\citep{huang2025benchmarkingllmsunittest}. While these pioneering benchmarks~\citep{zhuo2025bigcodebenchbenchmarkingcodegeneration,hai2025impactscontextsrepositorylevelcode,xu2025SWECompass} simulate realistic workflows, they predominantly rely on coarse-grained pass rates, thereby treating programming proficiency as a monolithic capability.

\begin{figure}[t]
    \centering
   \includegraphics[width=\linewidth]{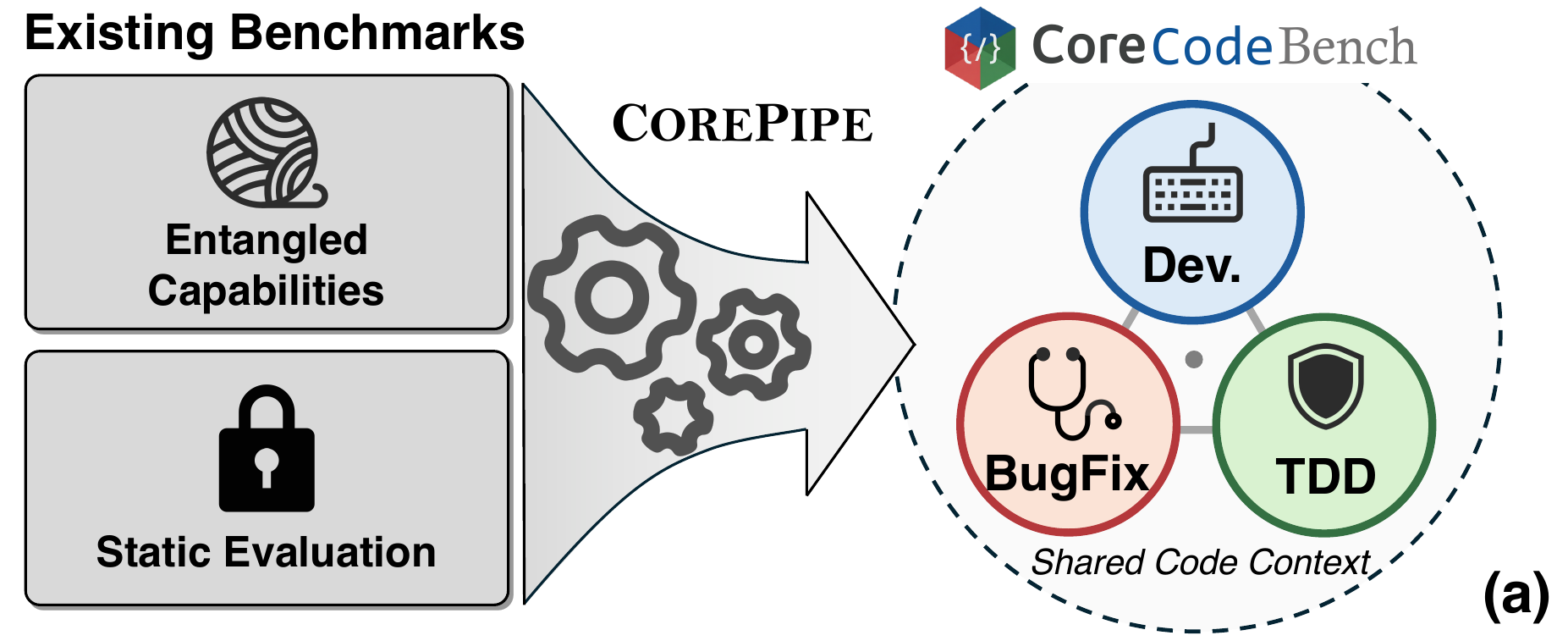}
    
     \vspace{2mm} 
     
     \includegraphics[width=\linewidth]{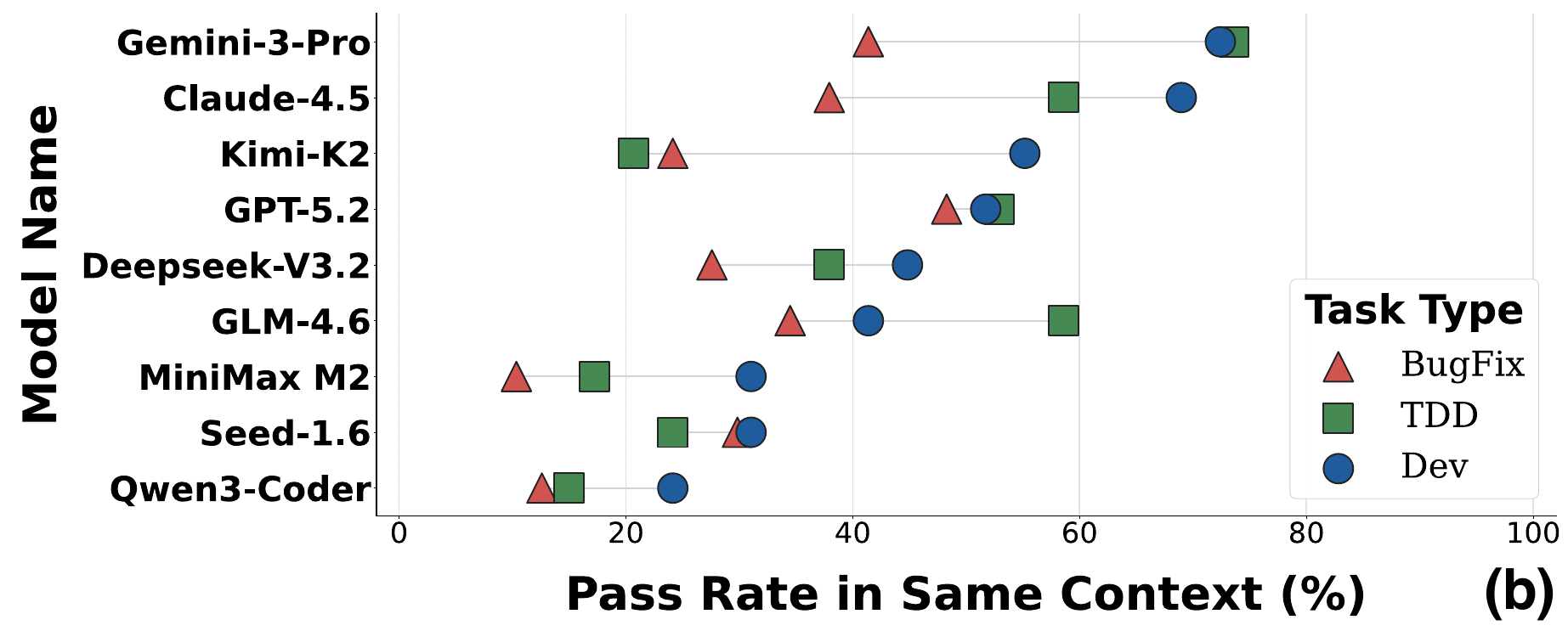}
    \caption{\textbf{Decomposing Code Intelligence.} (a) CoreCodeBench isolates distinct cognitive demands (Dev, BugFix, TDD) within an identical code context. (b) Performance comparison across these dimensions reveals significant \textbf{capability misalignment}, highlighting that coding proficiency is non-monolithic.}
    \label{fig:intro}
    \vspace{-15pt}
\end{figure}

However, this coarse-grained approach obscures the distinct cognitive demands inherent to different engineering scenarios. Real-world software engineering entails switching between generating code from intent (Development), {reasoning} about logic errors (BugFix), and {planning} verification constraints (Test Driven Development, TDD).  By conflating these dimensions under a single metric, existing benchmarks mask specific cognitive bottlenecks. As illustrated in Figure~\ref{fig:intro}(b), when evaluating models on these specialized tasks derived from an identical code context, we observe significant \textbf{capability misalignment}. Contrary to the expectation of uniform proficiency, models exhibit distinct ranking shifts across these dimensions. For instance, Kimi-K2 excels in Development but falters in BugFix and TDD, with a pass rate disparity exceeding 35\%. This dimensional inconsistency underscores that coding proficiency is not monolithic—high generative performance does not guarantee the grounded reasoning required for robust software engineering.

Compounding the granularity issue is the \textbf{static nature} of existing benchmarks, which limits their longevity and validity. By representing a fixed snapshot of the past, these benchmarks suffer from a dual vulnerability: data contamination, stemming from the memorization of open-source training data, and performance saturation, as rapidly evolving LLMs quickly master fixed difficulty levels. Consequently, the field urgently requires an automated, scalable pipeline capable of continuously transforming code contexts into diverse tasks. Such a framework must support dynamic difficulty scaling to prevent saturation and employ novel transformations to mitigate memorization, ensuring evaluation remains a rigorous, moving target aligned with the frontiers of code intelligence.

To address these challenges, we introduce \textsc{CoreCodeBench}, a \textbf{Co}nfigurable \textbf{Re}pository-level Benchmark designed to dissect LLM coding capabilities through atomized tasks. Leveraging the novel \textsc{CorePipe} framework, we automatically extract and transform 12 diverse Python repositories into a comprehensive suite of 1,524 evaluation tasks. Our approach offers three key advantages: (1) \textbf{Multi-dimensional Capability Isolation}: By generating six distinct task types on the same code context, we explicitly quantify performance disparities across distinct cognitive demands, distinguishing between generation, reasoning, and planning.  (2) \textbf{Controllable Difficulty}: We dynamically modulate task complexity by manipulating factors such as mask length and dependency depth, thereby mitigating performance saturation and ensuring sustainable evaluation. (3) \textbf{Superior Data Quality}: Our automated pipeline \textsc{CorePipe} ensures high reliability without manual intervention. CoreCodeBench achieves a 78.55\% validity yield, significantly surpassing the 31.7\% retention rate of manually filtered benchmarks like SWE-bench-Verified~\citep{openai2024swebench}.

To validate our framework, we conduct extensive experiments on a wide range of state-of-the-art (SoTA) LLMs. Our analysis probes the underlying nature of code intelligence, revealing a significant performance misalignment across distinct cognitive demands. Crucially, the distinct performance rankings observed here compared to other benchmarks suggest that our taxonomy captures distinct dimensions of capability overlooked by monolithic evaluations. Furthermore, we demonstrate the benchmark's sustainable scalability through graded difficulty levels, verifying that our complexity adjustments effectively mitigate performance saturation and maintain rigorous challenges for evolving models. Finally, the reliability of our automated pipeline is corroborated by fine-tuning experiments, where the successful acquisition of capabilities by smaller models confirms that CoreCodeBench generates canonical and valid supervision. In summary, our key contributions are as follows:
\begin{itemize}[leftmargin=10pt]
\item We propose a fine-grained taxonomy that disentangles programming proficiency into different tasks with distinct cognitive demands, moving beyond monolithic metrics to diagnose specific cognitive bottlenecks in LLMs.

\item We introduce CoreCodeBench, a contamination-resilient benchmark constructed via \textsc{CorePipe}. Beyond achieving a 78.55\% validity yield, it ensures sustainable evaluation by dynamically scaling difficulty to mitigate performance saturation.

\item We uncover a critical capability misalignment by systematically analyzing the inter-dependencies between distinct cognitive demands, thereby elucidating the internal structure of code intelligence and mapping how these dimensions correlate and diverge.

\end{itemize}

\begin{figure*}[t]
    \centering
    \includegraphics[width=\linewidth]{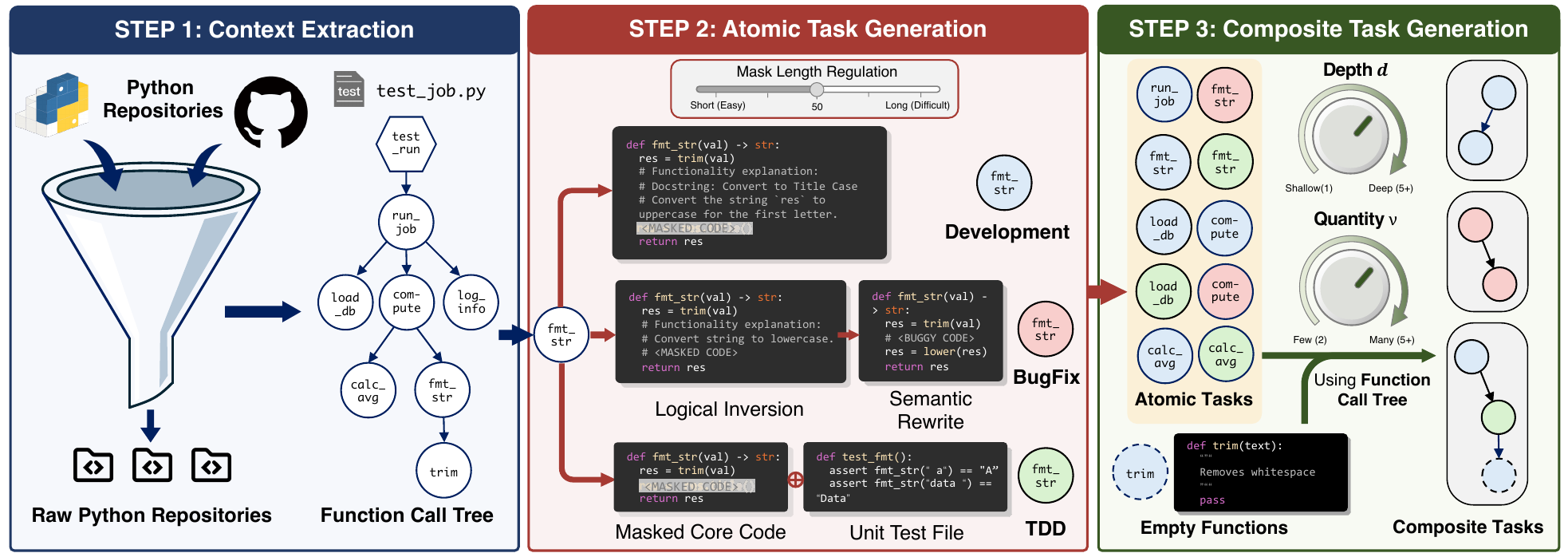}
    \caption{\textbf{Overview of the \textsc{CorePipe} Framework.} (i) \textit{Context Extraction} builds verifiable Function Call Trees from unit tests. (ii) \textit{Atomic Task Generation} isolates cognitive demands (Dev., BugFix, TDD) within identical contexts. (iii) \textit{Composite Task Scaling} aggregates atomic tasks into subgraphs, modulating difficulty via Dependency Depth ($d$) and Task Quantity ($\nu$) to prevent saturation.}
    \vspace{-8pt}
\end{figure*}

\section{\textsc{CorePipe} Framework}
To construct CoreCodeBench, we develop \textsc{CorePipe}, an automated framework designed to transform static repositories into dynamic, atomized evaluation tasks. As illustrated in Figure~\ref{fig:pipeline}, the framework proceeds through three stages: context extraction, atomic task generation with cognitive isolation, and composite task scaling. This pipeline guarantees rigorous verifiability while enabling dynamic difficulty scaling.

\subsection{Repository Context Extraction}
We select 12 open-source Python repositories from \href{https://pypi.org/}{PyPI} (detailed in Appendix~\ref{app:repo}) based on three rigorous criteria: (1) \textit{Activeness}: maintained within the last six months; (2) \textit{Test Coverage}: test files constitute $>15\%$ of the codebase; and (3) \textit{Complexity}: $>5,000$ LOC with cross-module dependencies.

To operationalize these repositories, we establish a precise mapping between source implementation and unit tests. We employ a hybrid approach combining LLM-based structural analysis with automated rule-based matching to generate \texttt{<source, test>} pairs. Subsequently, we perform dynamic tracing using a customized \texttt{pycallgraph}~\citep{pycallgraph} to construct a \textbf{Function Call Tree}. This tree structure, where nodes represent functions and edges represent invocation dependencies, serves as the backbone for both identifying core logic and composing complex tasks.

\subsection{Atomic Task Generation}
We generate atomic tasks by isolating single-function logic into distinct engineering scenarios. To ensure resilience against data contamination, our pipeline applies fine-grained transformations that alter the code's surface form and structural context. This prevents LLMs from simply recalling exact solutions from pre-training data, compelling them to reason grounded in the modified context.

\paragraph{Core Code Identification.}
To ensure evaluation substance, we filter out trivial utilities and focus on \textbf{core functions}. We prompt LLMs to identify semantically central AST blocks, filtering out trivial utilities. Crucially, we modulate the difficulty of the resulting tasks by varying the mask length of these blocks. We then validate the centrality of these blocks by executing unit tests on the masked code; only functions where masking triggers test failures are retained, ensuring the missing logic is essential for correctness. Manual inspection of 50 random samples confirms {100\% accuracy} in capturing essential logic, validating the reliability of this automated approach.

\paragraph{Development.}
 We mask the identified core code blocks and employ GPT-4o~\citep{openai2024helloGPT4o} to generate structured functional descriptions. Crucially, our ablation study confirms that model rankings are consistent across four distinct generator backbones (Spearman's $\rho = 1.0$, see Appendix~\ref{app:model-selection}). This validates that our metrics reflect intrinsic model capabilities, independent of generator-specific stylistic biases. To ensure quality, we introduce Claude-3.5-Sonnet~\citep{anthropic2024claude35} as a critic model to provide feedback. The generator refines the description based on this feedback. Finally, we apply an Information Gain (IG) filter (detailed in Appendix~\ref{app:igeval}) to discard samples where the generated explanation fails to provide effective guidance compared to the code context alone. 

\paragraph{BugFix.}
To evaluate the capability to diagnose and rectify logic errors, we extend the Development tasks by replacing the correct implementation with a buggy version. 
Unlike simple syntactic perturbations~\citep{liu2025mdevalmassivelymultilingualcode} or adversarial injections targeting model blind spots~\citep{ibrahimzada2025challengingbugpredictionrepair}, which often lack semantic coherence or human-like intent, we employ a cascaded logic-implementation synthesis. Specifically, we prompt advanced models (e.g., GPT-4o) to design complex logical fallacies (e.g., boundary neglect, state mismanagement), and subsequently leverage smaller models (e.g., Qwen3-Coder30B) to implement these flawed designs. 
This two-stage approach leverages the reasoning depth of large models to ensure logical complexity, while utilizing the stochasticity of smaller models to introduce natural implementation noise. Consequently, this bypasses the bias of LLMs to generate bug, yielding bugs that manifest as realistic developer slips rather than artificial artifacts (more details and case studies in Appendix~\ref{app:bugfix-casestudy}).

\paragraph{Test-Driven Development (TDD).}
TDD task evaluate the ability to implement logic strictly based on verification constraints~\citep{Mathews_2024,ahmed2024tddbenchverifiedllmsgenerate}. We directly utilize the core code blocks identified previously, masking the implementation while providing the corresponding unit tests as the specification. This paradigm is particularly relevant to modern LLM-assisted programming workflows, where models must align generated code with rigid pre-defined tests to ensure reliability.

\subsection{Composite Task Generation}\label{multi-func-gen}
To bridge the gap between isolated atomic tasks and real-world engineering, we introduce Composite Tasks (instantiated as \textit{multi-function problems}), which require models to reason over a subgraph of the function call tree rather than isolated nodes.

We define a composite task as a subgraph of the call tree containing $n$ atomic problems. 
To simulate realistic scenarios where primary logic relies on supporting components, we introduce \textbf{Empty-Function} nodes specifically for Dev. and TDD tasks. These nodes are integral parts of the subgraph but are stripped to their function signatures, representing auxiliary utilities that must be implemented synchronously to support the main objectives. Crucially, to prevent under-specification, we verify that the execution trace of the provided tests traverses these nodes, ensuring the main task strictly depends on their implementation. Furthermore, we enforce strict connectivity constraints during subgraph sampling to ensure that all selected nodes form a valid, executable dependency chain (detailed generation algorithms in Appendix~\ref{app:multi-problem-rules}).

Crucially, the complexity of composite tasks is explicitly controlled via two hyper-parameters to mitigate performance saturation: 
(1) \textbf{Dependency Depth ($d$)}, which regulates the scope of context by limiting the maximum depth of the call chain; and 
(2) \textbf{Task Quantity ($\nu$)}, which defines the total number of functions ($2\leq n \leq \nu$) requiring simultaneous modification.
By systematically adjusting $d$ and $\nu$, we create a graded difficulty scale that ensures the benchmark remains challenging for rapidly evolving models.

\begin{table}[t]
    \caption{\textbf{Statistics of CoreCodeBench}. \# Func and \# Lines denote the average number of functions and gold solution lines, respectively.}
    \label{tab:datastatistic}
    \centering
    \resizebox{0.9\linewidth}{!}{%
    \small
    \setlength{\tabcolsep}{3.5pt} 
    \renewcommand{\arraystretch}{1}
    \begin{tabular}{llccc}
    \toprule
    \textbf{Category} & \textbf{Task Type} & \textbf{\# Func} & \textbf{\# Lines} & \textbf{\# Prob} \\ 
    \midrule
    \multirow{3}{*}{Atomic} 
    & Development & 1.00 & 17.00 & 511 \\
    & BugFix & 1.00 & 38.00 & 315 \\         
    & TDD & 1.00 & 14.00 & 278 \\
    \midrule
    \multirow{4}{*}{Composite} 
    & Multi-Dev & 3.85 & 53.92 & 167 \\
    & Multi-BugFix & 2.00 & 62.34 & 10 \\         
    & Multi-TDD & 4.07 & 67.30 & 152 \\ \cmidrule{2-5}
    & Difficult & 4.75 & 65.66 & 91 \\
    \bottomrule
    \end{tabular}
    }
    \vspace{-2mm}
\end{table}

\section{CoreCodeBench Dataset}

\paragraph{Benchmark Composition.}
CoreCodeBench comprises 1,524 problems from 12 Python repositories. 
As shown in Table~\ref{tab:datastatistic}, tasks are categorized into Atomic (single-function) and Composite (orchestrating interdependent functions via call-tree subgraphs) levels across Development, BugFix, and TDD types.
Notably, we retain the \textit{Multi-BugFix} task despite its limited size (10) to preserve taxonomic symmetry, acknowledging the high difficulty of synthesizing valid interdependent bugs.
Additionally, we introduce a \textit{Difficult} subset (task quantity $\nu=\infty$) to probe the reasoning ceiling of current models.
Dataset details, prompts, and robustness checks are provided in Appendix~\ref{app:datasource}, \ref{app:problemprompt}, and \ref{app:promptvariation}.

\paragraph{Evaluation Protocol.}
We evaluate performance using unit tests. Crucially, we perform a pre-generation execution (denoted as \textit{retest}) on the initial masked or buggy code. This step records $N_{\text{retest}}$, the number of tests that pass \textit{without} any model modification.
We employ two metrics:
(1) \textbf{AC@1}~\citep{chen2021evaluatinglargelanguagemodels}: The percentage of problems where the generated solution passes all unit tests.
(2) \textbf{AC Rate}: A fine-grained metric measuring the proportion of fixed test cases:
\begin{equation}
\small
\text{AC Rate} = \frac{N_{\text{pass}} - N_{\text{retest}}}{N_{\text{total}} - N_{\text{retest}}},
\end{equation}
where $N_{\text{pass}}$ is the number of passed tests after generation. By subtracting $N_{\text{retest}}$, this metric strictly measures the model's contribution to fixing previously failing tests.

\section{Experiments}
We conduct a comprehensive evaluation to validate the quality of CoreCodeBench and probe the code intelligence of SoTA LLMs. Our analysis addresses four research questions:
\begin{itemize}[leftmargin=10pt]
    \item \textbf{RQ1}: Does the automated \textsc{CorePipe} generate high-quality, reliable evaluation tasks?
    \item  \textbf{RQ2}: Do LLMs exhibit consistent proficiency across distinct cognitive demands?
    \item \textbf{RQ3}: Can CoreCodeBench effectively mitigate performance saturation and sustain challenge through dynamic complexity modulation?
    \item \textbf{RQ4}: Does our fine-grained taxonomy capture dimensions of capability overlooked by existing monolithic benchmarks?
\end{itemize}

\subsection{Experimental Setups}\label{sec:setup}
We evaluate a comprehensive suite of 9 SoTA LLMs.
Our selection includes leading proprietary models, including {GPT-5.2}~\citep{gpt52}, {Claude-4.5-Opus}~\citep{claude45opus}, {Gemini-3-Pro}~\citep{gemini3}, and {Doubao-Seed1.6}~\citep{seed16}, as well as high-performance open-weights models, including {Qwen3-Coder-480B-A3B-Instruct}~\citep{qwencoder}, GLM-4.6~\citep{glm46}, Kimi-K2~\citep{kimik2}, MiniMax-M2~\citep{minimaxm2} and {DeepSeek V3.2}~\citep{deepseekv32}. For brevity, we hereafter refer to these models using only their family names and versions. Notably, for the correlation analyses in Section~\ref{subsec:rq4_validity}, we extend this set with 3 additional LLMs to enhance statistical reliability. All evaluations are performed in a zero-shot setting using greedy decoding, detailed implementation settings is in Appendix~\ref{app:implementationDetails}.



\subsection{Reliability of \textsc{CorePipe} (RQ1)}\label{subsec:rq1_quality}
\paragraph{Human Inspection.} We conduct a large-scale manual inspection on 360 development problems, covering $360/511=\mathbf{70.5\%}$ of the single-dev tasks.
Experienced engineers evaluate the problems based on readability, accuracy, and completeness (criteria detailed in Appendix~\ref{app:humanAnno}). This inspection yields a \textbf{78.55\%} qualification rate (with inter-annotator agreement rate > 95\%), significantly surpassing the 31.7\% retention rate of manually curated benchmarks like SWE-bench-Verified~\citep{openai2024swebench}. This result confirms that \textsc{CorePipe}'s automated mechanisms produces high-fidelity tasks suitable for reliable evaluation. For users requiring maximum precision, we also release the manually verified subset as \texttt{CoreCodeBench-Dev-Verified}. Note that TDD, BugFix and multi-function tasks are derived deterministically from existing code and tests without LLM-generated context, ensuring inherent validity without the need for manual text verification.

\begin{table}[t]
\caption{\textbf{Leaderboard on Atomic Tasks.} Rate indicates AC Rate (\%). Best in \textbf{bold}, second \underline{underlined}.}
\label{tab:SingleFunction}
\centering
\scriptsize
\setlength{\tabcolsep}{2.5pt} 
\renewcommand{\arraystretch}{1.0}
\resizebox{\linewidth}{!}{%
\begin{tabular}{clcccccc}
\toprule
\multirow{2}{*}{\textbf{Type}} & \multirow{2}{*}{\textbf{Model}} & \multicolumn{2}{c}{\textbf{Dev.}} & \multicolumn{2}{c}{\textbf{BugFix}} & \multicolumn{2}{c}{\textbf{TDD}} \\ 
\cmidrule(lr){3-4} \cmidrule(lr){5-6} \cmidrule(lr){7-8}
& & \textbf{AC@1} & \textbf{Rate} & \textbf{AC@1} & \textbf{Rate} & \textbf{AC@1} & \textbf{Rate} \\ 
\midrule
\multirow{4}{*}{\rotatebox{90}{API}} 
& Gemini-3-Pro & \underline{67.24} & \textbf{87.72} & \textbf{62.77} & \textbf{78.58} & \textbf{73.39} & \textbf{89.33} \\
& Claude-4.5 & \textbf{68.26} & 86.56 & 51.66 & 63.25 & 61.24 & 83.92 \\
& GPT-5.2 & 57.66 & 81.80 & \underline{53.56} & \underline{70.92} & \underline{67.35} & 83.27 \\
& Seed-1.6 & 39.70 & 69.35 & 41.57 & 64.70 & 37.93 & 52.90 \\
\midrule
\multirow{5}{*}{\rotatebox{90}{Open-Source}} 
& Kimi-K2 & 58.95 & 86.34 & 29.57 & 45.63 & 50.99 & 79.45 \\
& DeepSeek-V3.2 & 62.06 & \underline{87.16} & 41.00 & 60.83 & 57.59 & 83.04 \\
& Qwen3-Coder & 57.35 & 83.96 & 36.46 & 55.03 & 54.19 & 83.96 \\
& GLM-4.6 & 52.56 & 83.26 & 42.46 & 63.68 & 65.23 & \underline{87.33} \\
& MiniMax-M2 & 23.12 & 53.50 & 18.56 & 30.84 & 33.17 & 61.96 \\
\bottomrule
\end{tabular}
}
\end{table}

\begin{table}[t]
\caption{\textbf{Leaderboard on Composite Tasks.} Rate indicates AC Rate (\%). Best in \textbf{bold}, second \underline{underlined}.}
\label{tab:MultiFunction}
\centering
\scriptsize
\setlength{\tabcolsep}{2.5pt} 
\renewcommand{\arraystretch}{1.0}
\resizebox{\linewidth}{!}{%
\begin{tabular}{clcccccc}
\toprule
\multirow{2}{*}{\textbf{Type}} & \multirow{2}{*}{\textbf{Model}} & \multicolumn{2}{c}{\textbf{Dev.}} & \multicolumn{2}{c}{\textbf{BugFix}} & \multicolumn{2}{c}{\textbf{TDD}} \\ 
\cmidrule(lr){3-4} \cmidrule(lr){5-6} \cmidrule(lr){7-8}
& & \textbf{AC@1} & \textbf{Rate} & \textbf{AC@1} & \textbf{Rate} & \textbf{AC@1} & \textbf{Rate} \\ 
\midrule
\multirow{4}{*}{\rotatebox{90}{API}} 
& Gemini-3-Pro 
    & \textbf{18.44} & \textbf{35.08} 
    & 0.00 & \underline{15.99} 
    & \textbf{20.67} & \textbf{42.16} \\
& Claude-4.5
    & \underline{16.06} & 27.09
    & 0.00 & 13.85 
    & \underline{17.80} & 25.51 \\
& GPT-5.2 
    & 3.82 & 9.34 
    & 0.00 & 13.85 
    & 13.76 & 20.34 \\
& Seed-1.6 
    & 0.30 & 7.34 
    & 0.00 & 13.85 
    & 7.61 & 20.87 \\
\midrule
\multirow{5}{*}{\rotatebox{90}{Open-Source}} 
& Kimi-K2 
    & 8.78 & 25.06 
    & 0.00 & 2.14 
    & 5.63 & 16.21 \\
& DeepSeek-V3.2 
    & 15.36 & \underline{32.48}
    & 0.00 & 13.85 
    & 9.98 & 27.80 \\
& Qwen3-Coder
    & 12.80 & 25.89 
    & 0.00 & \textbf{18.66} 
    & 8.77 & 21.01 \\
& GLM-4.6 
    & 6.76 & 24.77 
    & 0.00 & \underline{15.99} 
    & 15.91 & \underline{31.32} \\
& MiniMax-M2 
    & 0.77 & 5.30 
    & 0.00 & 0.00 
    & 2.49 & 10.99 \\
\bottomrule
\end{tabular}
}
\end{table}

\paragraph{Fine-tuning Validation.}
To further corroborate that CoreCodeBench provides canonical supervision, we perform a controlled fine-tuning experiment. We adopt a repository-level split (11 for training, 1 for testing) to prevent data leakage. We fine-tune Qwen3-8B~\citep{qwen2025qwen3} on the training split and evaluate their performance on the unseen test repositories. Training details and experiment results are available in Appendix~\ref{app:ft_exp}. The result yields significant improvements on 5 metrics across all three atomic tasks (e.g., Dev. AC@1 increases by 19.15\%), confirming that CoreCodeBench serves as a canonical supervision signal.

\subsection{Main Results: Overall Performance}
\label{subsec:main_results}

Table~\ref{tab:SingleFunction} reports the performance on atomic tasks, while Table~\ref{tab:MultiFunction} summarizes the results on composite tasks.
Confidence intervals and results for additional LLMs are provided in Appendix~\ref{app:evaluation-results}.

\paragraph{Atomic Tasks.}
As shown in Table~\ref{tab:SingleFunction}, proprietary models establish a clear lead, with {Gemini-3-Pro} and {Claude-4.5-Opus} achieving top-tier performance.
While open-source models like GLM-4.6 and Qwen3-Coder have narrowed the gap in generation-heavy tasks (Development and TDD), performance drops significantly on {BugFix} across the board.
This decline is particularly pronounced for open-source models, where AC@1 scores often halve compared to Development.
This suggests that diagnosing logic errors remains a distinct bottleneck compared to generation, a phenomenon we analyze further in Section~\ref{subsec:rq2_misalignment}.

\begin{figure}[t]
    \centering
    \includegraphics[width=0.8\linewidth]{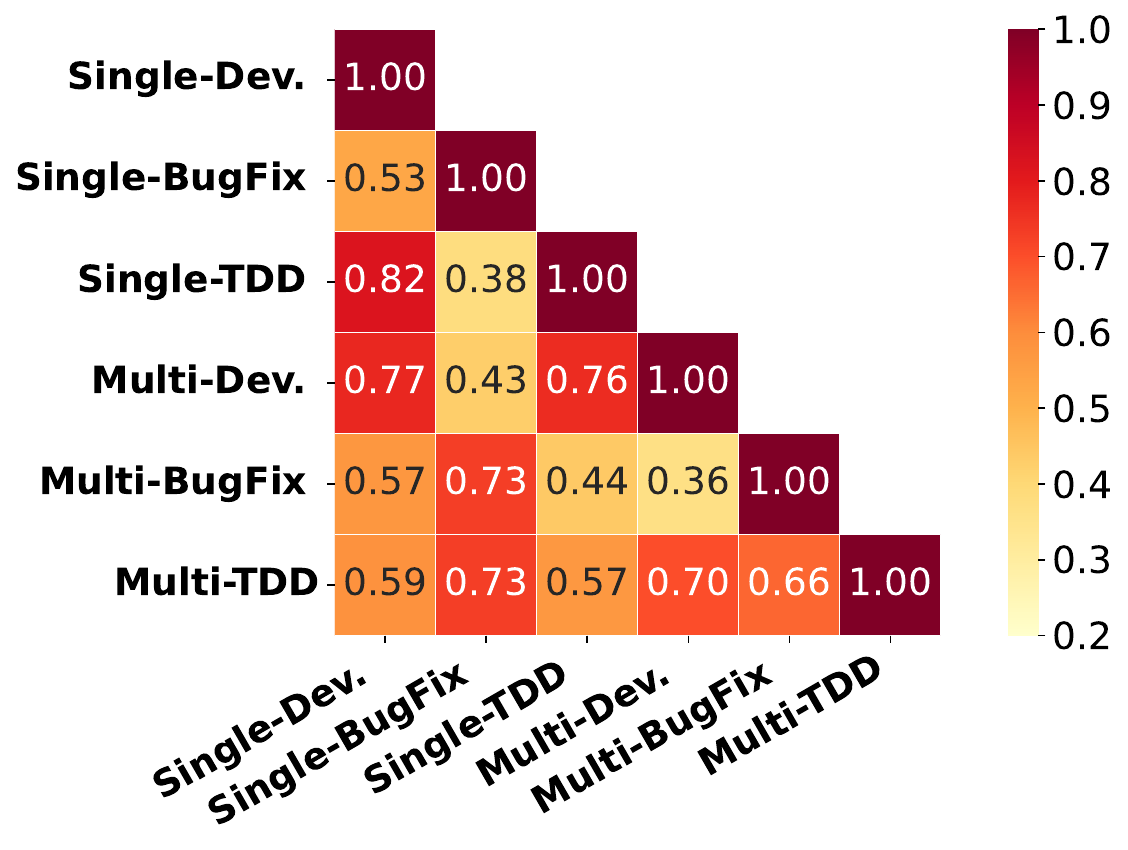}
       \caption{Pearson correlation matrix across six tasks.}
        \label{fig:corr_matrix}
        \vspace{-5pt}
\end{figure}

\paragraph{Composite Tasks.}
Table~\ref{tab:MultiFunction} reveals a dramatic increase in difficulty as we move to repo-level composite tasks.
Most strikingly, {Multi-BugFix} proves to be an extremely formidable challenge, with all models failing to solve a single instance (0.00\% AC@1), though non-zero AC Rates indicate partial progress.
This performance collapse correlates with a critical behavioral observation: in multi-function scenarios, models are required to complete multiple functions within a single response.
Ideally, LLM would demonstrate planning by ordering implementations based on dependencies (e.g., implementing utility functions before their callers).
However, all models strictly adhered to the prompt's sequence, ignoring dependency logic. 
This rigid linearity exposes a critical deficit in hierarchical reasoning, rendering current models ineffective in complex, interdependent environments.

\subsection{Capability Misalignment (RQ2)}\label{subsec:rq2_misalignment}
\paragraph{Task Correlation Analysis.}
To quantify the relationships between different evaluation dimensions, we analyze the performance alignment across the six task types. Specifically, we construct performance vectors for each task using the AC Rate scores of all evaluated models. We then calculate the Pearson correlation coefficient~\citep{pearson1896regression} between every pair of task vectors, as visualized in Figure~\ref{fig:corr_matrix}. Using AC Rate ensures meaningful correlation analysis even for challenging tasks like Multi-BugFix where AC@1 scores are sparse.

\begin{figure}[t]
     \centering
        \includegraphics[width=0.95\linewidth]{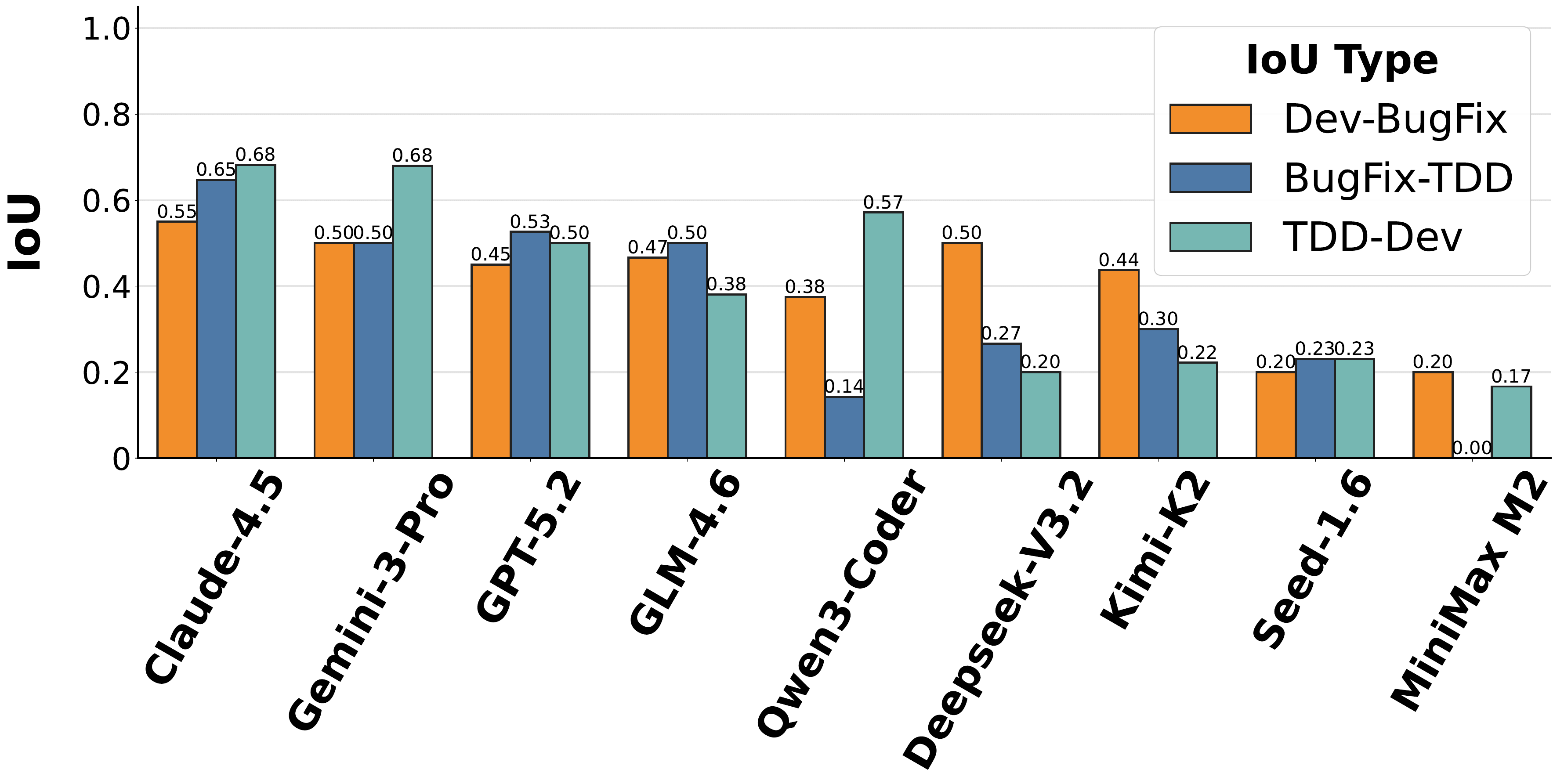}
        \caption{Quantifying Capability Misalignment via Inter-task IoU. The generally low IoU values reveal significant inconsistency between tasks.}
        \label{fig:iou_consistency}
        \vspace{-5pt}
\end{figure}

\begin{figure*}[t]
    \centering
      \captionsetup[subfigure]{labelformat=parens}
    \begin{subfigure}[b]{0.45\textwidth}
        \centering
        \includegraphics[width=\linewidth]{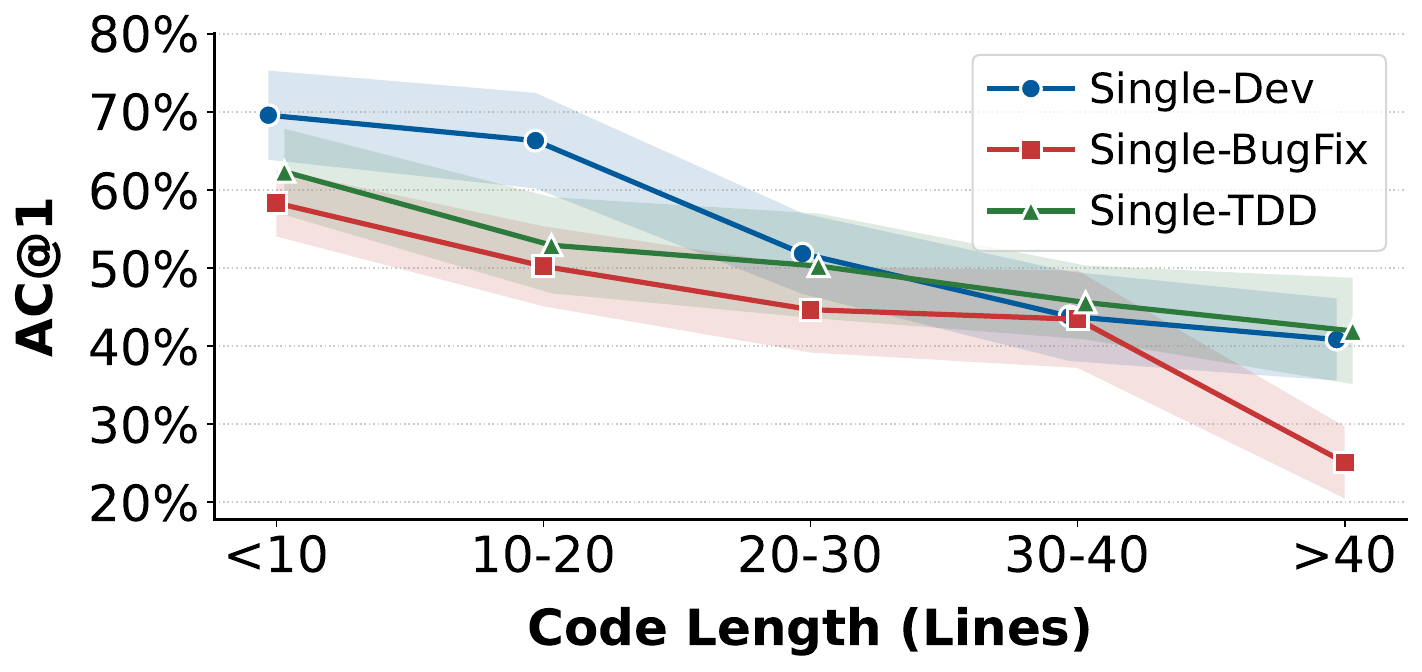}
        \caption{Impact of Code Length.}
    \end{subfigure}
    \hfill
    \begin{subfigure}[b]{0.45\textwidth}
        \centering
        \includegraphics[width=\linewidth]{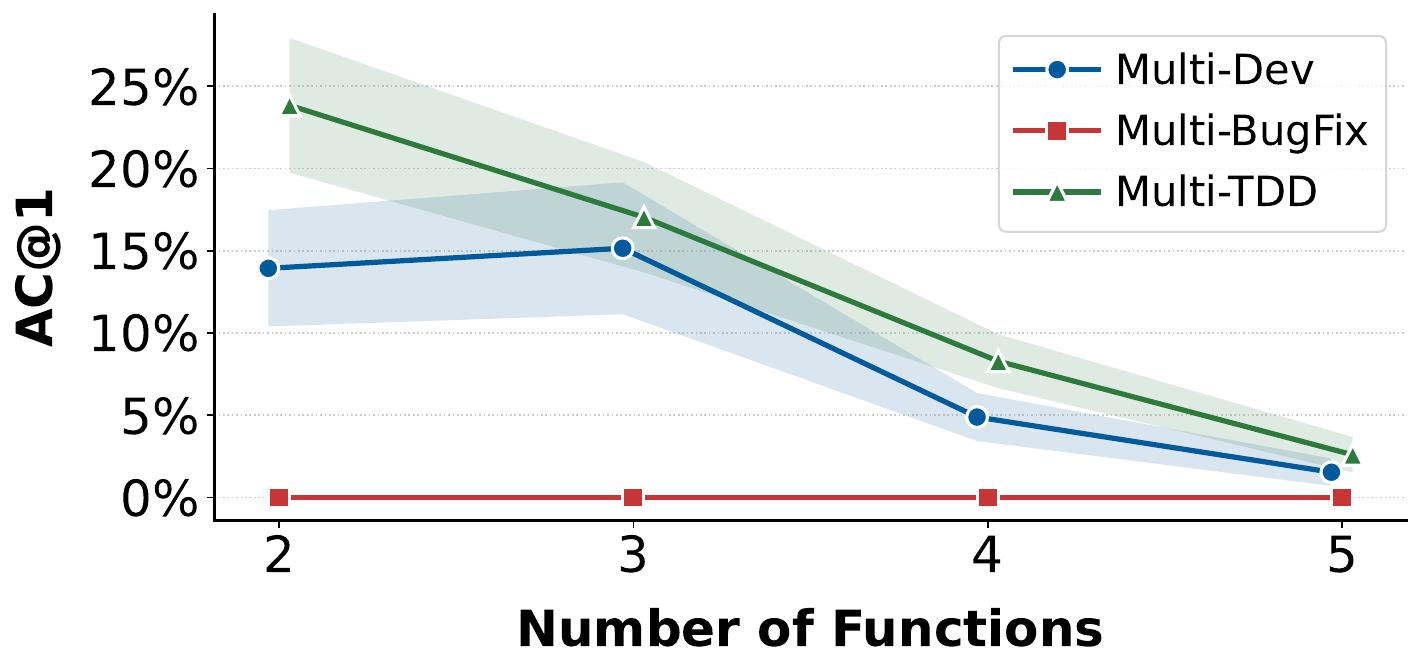}
        \caption{Impact of Task Quantity $\nu$.}
    \end{subfigure}
    \vspace{-5pt}
    \caption{\textbf{Difficulty Scaling Analysis.} (a) Performance consistently declines as the code length increases. (b) Increasing the number of interdependent functions ($\nu$) triggers a performance collapse.}
  \label{fig:difficulty_scaling}
  \vspace{-5pt}
\end{figure*}

\paragraph{Inter-task Consistency.} 
To further probe whether models possess a unified understanding of the code context, we analyze the consistency of their success across different tasks derived from the identical function using the Intersection-over-Union (IoU) metric. 
For any pair of tasks $A$ and $B$, IoU is calculated as: $\text{IoU}(A, B) = \frac{|S_A \cap S_B|}{|S_A \cup S_B|}$,
where $S_A$ and $S_B$ denote the sets of instances solved by the model in task $A$ and $B$, respectively.

As visualized in Figure~\ref{fig:iou_consistency}, the absolute IoU values reveal a pervasive lack of consistency.
Even for the strongest proprietary models (e.g., Claude-4.5, Gemini-3-Pro), the overlap is less than 0.7, implying that in at least 30\% of cases, proficiency is inconsistent across dimensions.
The landscape for open-source models is even more fragmented and diverse.
While Qwen3-Coder maintain moderate consistency between Development and TDD, their alignment with BugFix is drastically lower ($<0.2$).
This variance in IoU distributions confirms that different models rely on distinct, often disjoint capability profiles, lacking a robust, transferable mental model of the code.

\begin{figure}[t]
    \centering
    \includegraphics[width=\linewidth]{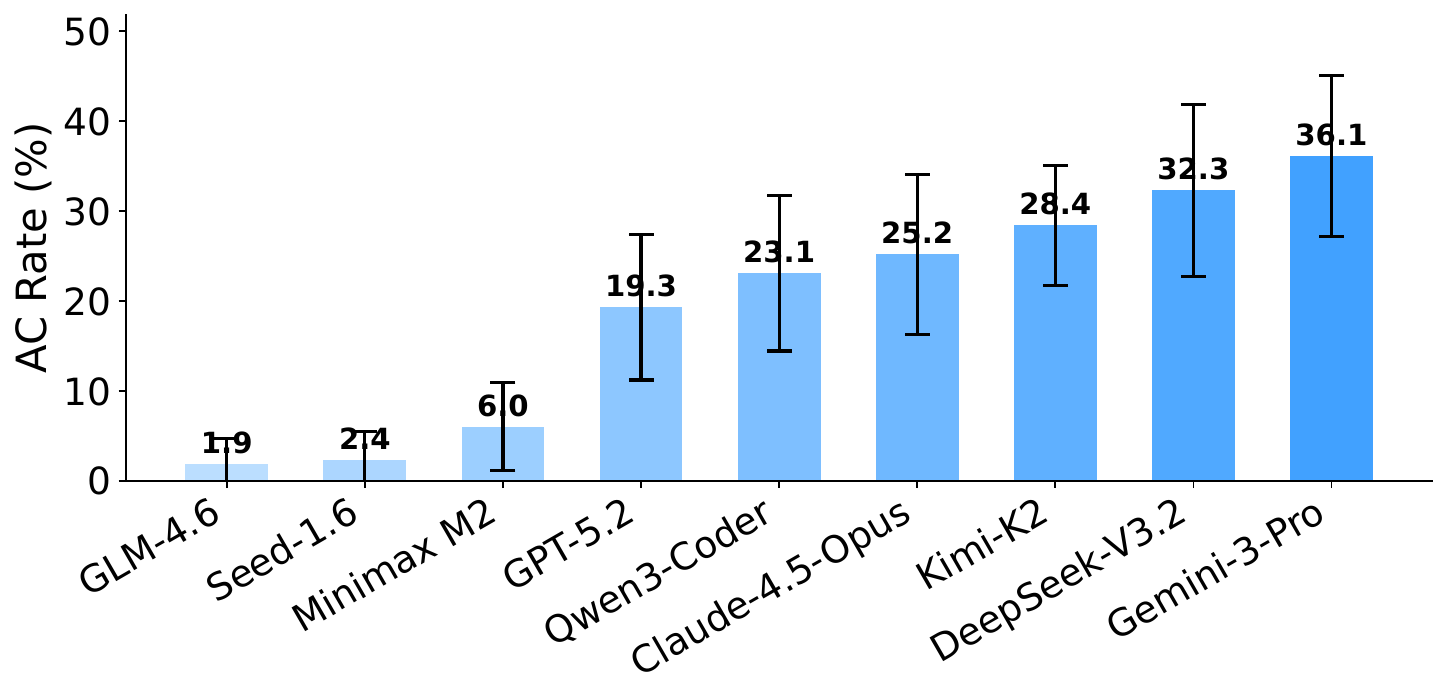}
    \vspace{-15pt}
    \caption{\textbf{Performance on CoreCodeBench-\textit{Difficult}}. } 
    \label{fig:difficult}
\vspace{-10pt}
\end{figure}

\subsection{Mitigating Performance Saturation (RQ3)}\label{subsec:rq3_difficulty}
A key design goal of CoreCodeBench is to prevent performance saturation by providing controllable difficulty gradients. We analyze how model performance scales with two key complexity factors: code length and dependency scope.

\paragraph{Impact of Mask Length.}
Figure~\ref{fig:difficulty_scaling}(a) illustrates the relationship between the length of the masked core code block (a configurable parameter in \textsc{CorePipe}) and model average Pass@1 across atomic tasks. We observe a consistent negative correlation: as the mask length increases, the average AC@1 drops significantly across all tasks. Notably, {BugFix} (Red line) exhibits the steepest decline in long contexts ($>30$ lines), falling below 25\%. This suggests that while models can manage short context repairs, their reasoning capability degrades rapidly as the search space for logic errors expands. Crucially, this trend confirms that by adjusting the mask length, \textsc{CorePipe} can effectively modulate the difficulty of atomic tasks, tailoring the challenge to match evolving model capabilities.

\paragraph{Impact of Task Quantity ($\nu$).}
As shown in Figure~\ref{fig:difficulty_scaling}(b), we further investigate the effect of dependency complexity by scaling the number of interdependent functions ($\nu$) in Multi-Function tasks.
Performance degrades sharply as $\nu$ increases, revealing a clear complexity cliff.
Notably, Multi-BugFix (Red line) remains near zero throughout. 
This is expected, as simultaneously diagnosing logic errors in just two interdependent functions ($\nu=2$) already exceeds the reasoning capacity of current models; increasing $\nu$ further only compounds this insurmountable challenge.
This confirms that manipulating $\nu$ serves as a potent lever to modulate difficulty, ensuring the benchmark remains challenging for future models.
To push this to the limit, we construct the \textbf{CoreCodeBench-Difficult} subset by setting $\nu=\infty$ (unbounded task quantity).
As shown in Figure~\ref{fig:difficult}, even the strongest models achieve AC Rates below 40\% on this subset.
This explicitly delineates the current ceiling of code intelligence, providing a rigorous testbed for driving future advancements.

\subsection{External Validity (RQ4)}
\label{subsec:rq4_validity}

\begin{figure*}[t]
    \centering
    \captionsetup[subfigure]{labelformat=parens}
    \begin{subfigure}[b]{0.47\textwidth}
        \centering
        \includegraphics[width=\linewidth]{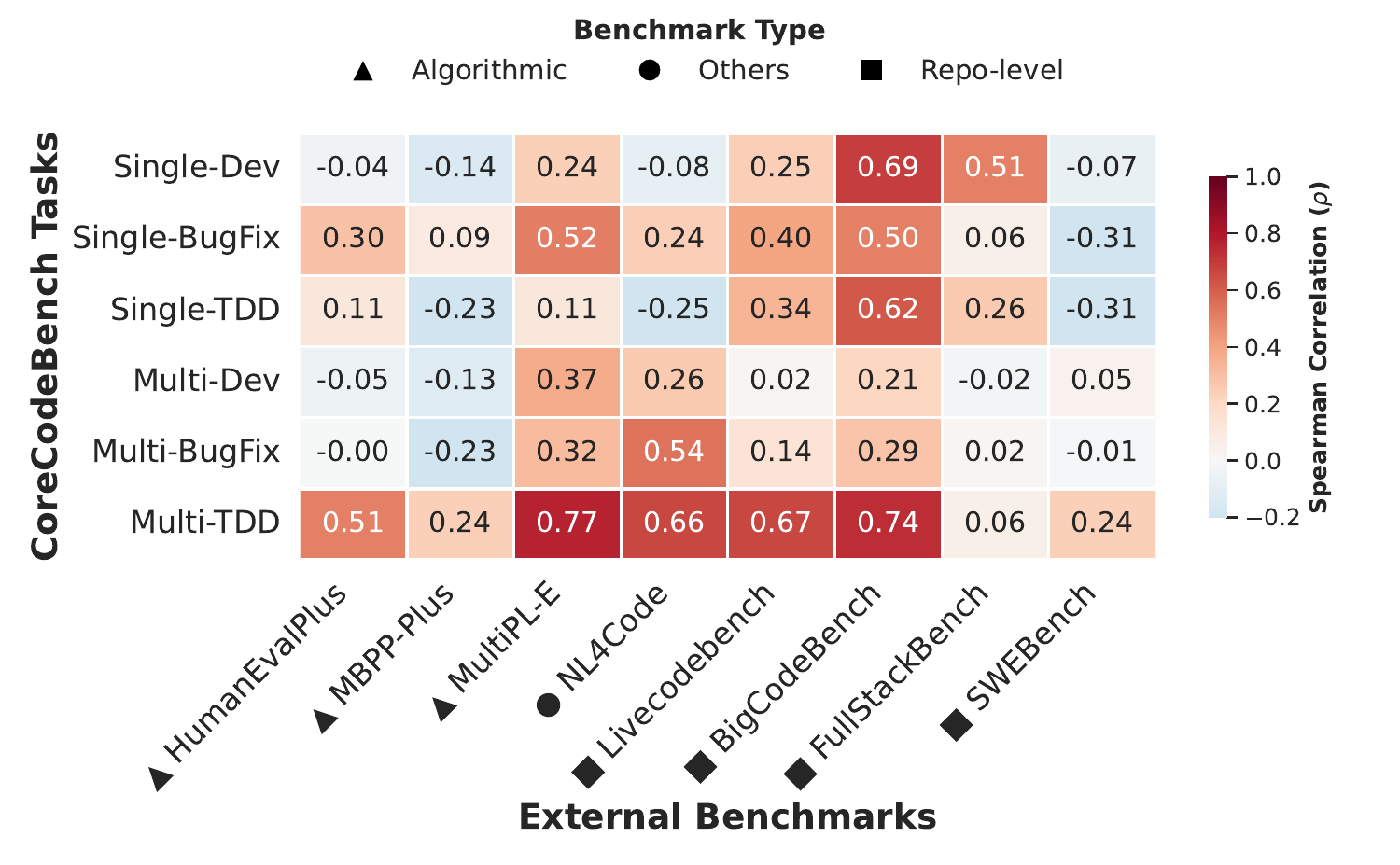}
        \caption{Correlation Heatmap. }
        \label{fig:coorelationheatmap}
    \end{subfigure}
    \hfill
    \begin{subfigure}[b]{0.47\textwidth}
        \centering
        \includegraphics[width=\linewidth]{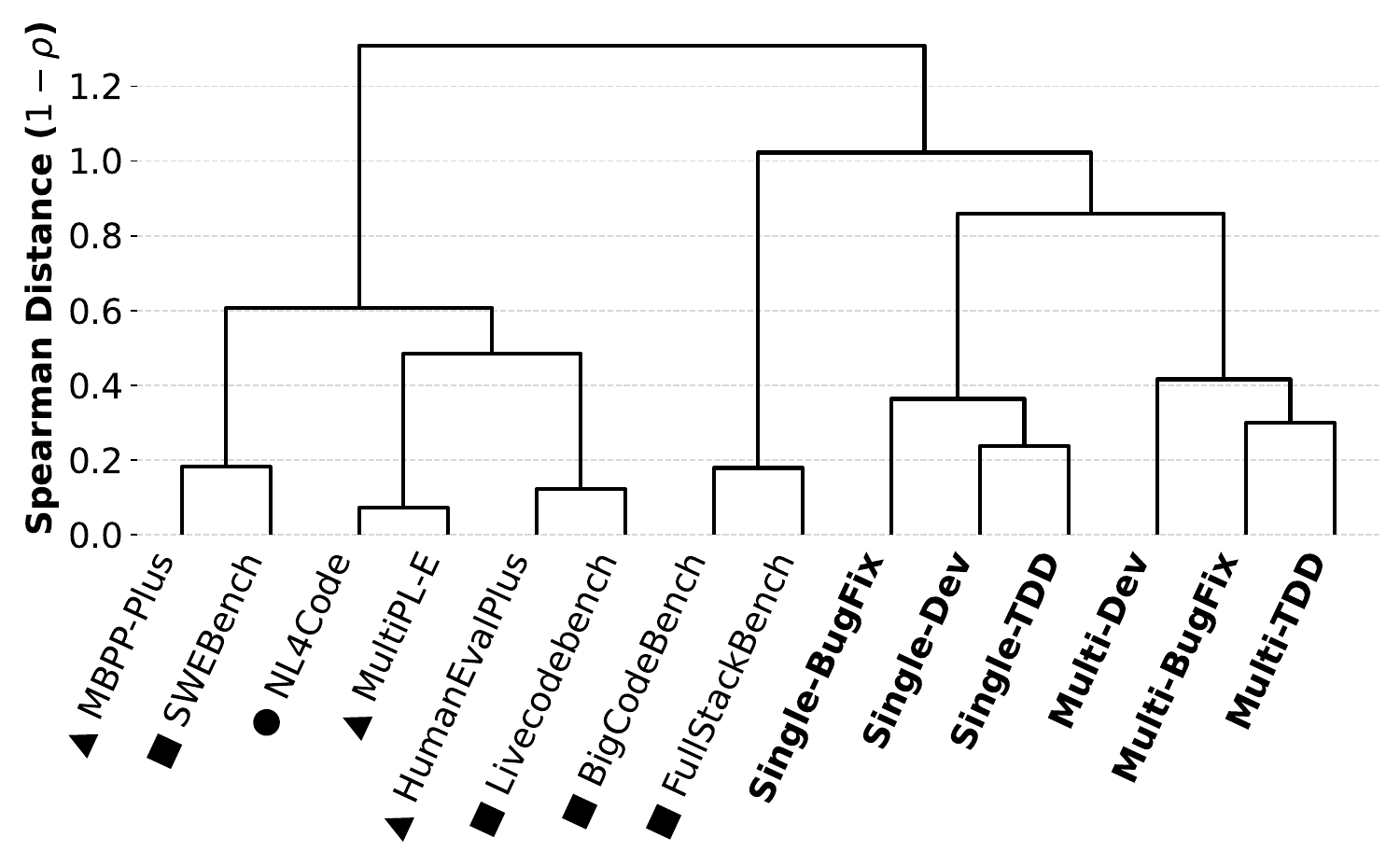}
        \caption{Hierarchical Clustering.}
        \label{fig:hierarchical_clustering}
    \end{subfigure}
     \vspace{-5pt}
    \caption{\textbf{Comparison with Existing Benchmarks.} (a) Heatmap shows limited alignment with external benchmarks. (b) Clustering reveals that CoreCodeBench forms a unique cluster distinct from current benchmarks.}
    \vspace{-4pt}
  \label{fig:external_validity}
\end{figure*}

To position CoreCodeBench within the broader evaluation ecosystem, we analyze its relationship with 8 major external benchmarks, ranging from algorithmic puzzles (including HumanEvalPlus~\citep{liu2023your}, MBPP-Plus~\citep{liu2023your} and MultiPL-E~\citep{cassano2022multiplescalableextensibleapproach}) to repository-level engineering tasks (including SWE-bench~\citep{openai2024swebench}, BigCodeBench~\citep{zhuo2025bigcodebenchbenchmarkingcodegeneration}, LiveCodeBench~\citep{jain2024livecodebench} and FullStackBench~\citep{cheng2024fullstack}).
We compute the Spearman correlation coefficients ($\rho$)~\citep{spearman1961proof} between the performance rankings of 12 LLMs (the 9 primary models plus 3 extended ones) across our six tasks and external benchmarks.

\paragraph{Correlation Landscape.}
The heatmap in Figure~\ref{fig:coorelationheatmap} visualizes the pairwise correlations, revealing the limited coverage of existing benchmarks.
Most external datasets correlate well with only specific subsets of our tasks, failing to capture the full spectrum of engineering capabilities.
For instance, algorithmic benchmarks align closely with our TDD tasks, because both paradigms rely on test-guided generation where models optimize code to satisfy rigid input-output constraints. In contrast, repository-level benchmarks exhibit a more mixed correlation pattern. This indicates that while existing benchmarks serve as specialized probes for monolithic proficiency, CoreCodeBench provides a more granular diagnosis, encompassing dimensions that are often conflated or overlooked.

\paragraph{Structural Uniqueness.}
The hierarchical clustering dendrogram in Figure~\ref{fig:hierarchical_clustering} further clarifies this distinction by revealing the latent taxonomy of the current benchmark landscape. In this visualization, the vertical height of the branches represents the dissimilarity between tasks; tasks that merge at lower heights are more strongly correlated. The analysis uncovers a fundamental bifurcation into two primary clusters. The traditional cluster (left) aggregates the majority of existing benchmarks, and CoreCodeBench forms a distinct branch (right). This structural separation serves as strong empirical evidence that CoreCodeBench introduces a novel evaluation perspective. Rather than merely replicating existing difficulty levels, it captures specific dimensions of engineering capability that differ significantly from the established evaluation ecosystem.

\section{Related Work}
\paragraph{Coding Capability of LLMs.}
LLMs have achieved remarkable proficiency in coding capabilities. On HumanEval~\citep{chen2021evaluatinglargelanguagemodels}, leading closed-source models like Claude-3.5-Sonnet~\citep{anthropic2024claude3} and GPT-4o~\citep{openai2024helloGPT4o}, as well as open-source DeepSeek-Coder-V2~\citep{deepseekai2024deepseekcoderv2breakingbarrierclosedsource} and Qwen2.5-Coder~\citep{hui2024qwen25codertechnicalreport}, all exceed 88\% AC@1. Similar mastery is observed on MBPP~\citep{austin2021programsynthesislargelanguage} with scores surpassing 85\%. Beyond function-level synthesis, the field has transitioned to complex software engineering; frontier models such as Claude 4.5 Opus~\citep{claude45opus}, Gemini 3 Pro~\citep{gemini3} reportedly achieve over 70\% resolution rates on SWE-bench~\citep{openai2024swebench}.

\paragraph{Repository-level Code Benchmarks.}
Existing repository-level benchmarks are typically designed for isolated scenarios, such as code completion~\citep{wu2025repomasterevalevaluatingcodecompletion, niu2023crosscodebenchbenchmarkingcrosstaskgeneralization}, functional code generation~\citep{hai2025impactscontextsrepositorylevelcode, yang2024execrepobenchmultilevelexecutablecode, fu2024codeapexbilingualprogrammingevaluation}, multi-stage software development~\citep{li2024promptinglargelanguagemodels}, or specialized engineering tasks like code translation~\citep{wang2025repotransbenchrealworldmultilingualbenchmark} and version migration ~\citep{liu2025migrationbenchrepositorylevelcodemigration}. Their underlying data is often derived from random masking~\citep{liu2023repobenchbenchmarkingrepositorylevelcode}, automated extraction of GitHub issues and pull requests ~\citep{jimenez2024swebenchlanguagemodelsresolve, pan2024codevbenchllmsunderstanddevelopercentric, li2025feabenchbenchmarkevaluatingrepositorylevel}, or manual curation~\citep{zhuo2025bigcodebenchbenchmarkingcodegeneration, li2024evocodebenchevolvingcodegeneration}, which renders the evaluation static, difficult to scale, and vulnerable to data contamination. Consequently, the field requires a dynamic framework to dissect the monolithic proficiency masked by static evaluations.

\section{Conclusion}
\label{sec:conclusion}
We introduce CoreCodeBench, a repository-level benchmark decomposing code intelligence into distinct cognitive dimensions. Leveraging the automated \textsc{CorePipe}, we transform repositories into 1,524 tasks with controllable difficulty and high quality without reliance on manual curation.
Our extensive evaluation reveals a significant capability misalignment: models exhibit uneven proficiency across development, debugging, and planning tasks, even within the identical code context.
This finding challenges the prevailing monolithic view of coding proficiency, demonstrating that robust software engineering requires the synergy of diverse, often imbalanced capabilities.
Furthermore, our structural analysis confirms that CoreCodeBench captures evaluation dimensions that differ from existing algorithmic and repair benchmarks.

\newpage
\section*{Limitations}
Despite the automated generation capabilities of \textsc{CorePipe}, our framework currently relies on the presence of high-coverage unit tests within source repositories. Consequently, repositories with sparse tests cannot be processed, potentially biasing the benchmark towards well-maintained projects. Future work will explore automated test generation to broaden this scope. Additionally, due to the stringent constraints required to synthesize valid interdependent logic errors, the \textit{Multi-BugFix} subset contains a limited number of instances (10). While valuable as a case study for extreme difficulty, its statistical power for ranking is inherently lower than other categories.
Finally, CoreCodeBench focuses exclusively on Python; extending support to other languages (e.g., Java, C++) remains a critical direction to evaluate cross-lingual engineering proficiency.




\bibliography{ref}

@misc{zhuo2025bigcodebenchbenchmarkingcodegeneration,
      title={BigCodeBench: Benchmarking Code Generation with Diverse Function Calls and Complex Instructions}, 
      author={Terry Yue Zhuo and Minh Chien Vu and Jenny Chim and Han Hu and Wenhao Yu and Ratnadira Widyasari and Imam Nur Bani Yusuf and Haolan Zhan and Junda He and Indraneil Paul and Simon Brunner and Chen Gong and Thong Hoang and Armel Randy Zebaze and Xiaoheng Hong and Wen-Ding Li and Jean Kaddour and Ming Xu and Zhihan Zhang and Prateek Yadav and Naman Jain and Alex Gu and Zhoujun Cheng and Jiawei Liu and Qian Liu and Zijian Wang and Binyuan Hui and Niklas Muennighoff and David Lo and Daniel Fried and Xiaoning Du and Harm de Vries and Leandro Von Werra},
      year={2025},
      eprint={2406.15877},
      archivePrefix={arXiv},
      primaryClass={cs.SE},
      url={https://arxiv.org/abs/2406.15877}, 
}

@misc{stepanov2015kendallcorrelationcoefficient,
      title={On the Kendall Correlation Coefficient}, 
      author={Alexei Stepanov},
      year={2015},
      eprint={1507.01427},
      archivePrefix={arXiv},
      primaryClass={math.ST},
      url={https://arxiv.org/abs/1507.01427}, 
}

@article{pycallgraph,
  title={Python call graph},
  author={Kaszuba, Gerald},
  journal={Internet: https://pycallgraph. readthedocs. io/en/master},
  year={2016}
}

@misc{fu2024codeapexbilingualprogrammingevaluation,
      title={CodeApex: A Bilingual Programming Evaluation Benchmark for Large Language Models}, 
      author={Lingyue Fu and Huacan Chai and Shuang Luo and Kounianhua Du and Weiming Zhang and Longteng Fan and Jiayi Lei and Renting Rui and Jianghao Lin and Yuchen Fang and Yifan Liu and Jingkuan Wang and Siyuan Qi and Kangning Zhang and Weinan Zhang and Yong Yu},
      year={2024},
      eprint={2309.01940},
      archivePrefix={arXiv},
      primaryClass={cs.CL},
      url={https://arxiv.org/abs/2309.01940}, 
}

@misc{liu2025mdevalmassivelymultilingualcode,
      title={MdEval: Massively Multilingual Code Debugging}, 
      author={Shukai Liu and Linzheng Chai and Jian Yang and Jiajun Shi and He Zhu and Liran Wang and Ke Jin and Wei Zhang and Hualei Zhu and Shuyue Guo and Tao Sun and Jiaheng Liu and Yunlong Duan and Yu Hao and Liqun Yang and Guanglin Niu and Ge Zhang and Zhoujun Li},
      year={2025},
      eprint={2411.02310},
      archivePrefix={arXiv},
      primaryClass={cs.CL},
      url={https://arxiv.org/abs/2411.02310}, 
}

@misc{jimenez2024swebenchlanguagemodelsresolve,
      title={SWE-bench: Can Language Models Resolve Real-World GitHub Issues?}, 
      author={Carlos E. Jimenez and John Yang and Alexander Wettig and Shunyu Yao and Kexin Pei and Ofir Press and Karthik Narasimhan},
      year={2024},
      eprint={2310.06770},
      archivePrefix={arXiv},
      primaryClass={cs.CL},
      url={https://arxiv.org/abs/2310.06770}, 
}

@misc{li2024evocodebenchevolvingcodegeneration,
      title={EvoCodeBench: An Evolving Code Generation Benchmark Aligned with Real-World Code Repositories}, 
      author={Jia Li and Ge Li and Xuanming Zhang and Yihong Dong and Zhi Jin},
      year={2024},
      eprint={2404.00599},
      archivePrefix={arXiv},
      primaryClass={cs.CL},
      url={https://arxiv.org/abs/2404.00599}, 
}

@misc{hui2024qwen25codertechnicalreport,
      title={Qwen2.5-Coder Technical Report}, 
      author={Binyuan Hui and Jian Yang and Zeyu Cui and Jiaxi Yang and Dayiheng Liu and Lei Zhang and Tianyu Liu and Jiajun Zhang and Bowen Yu and Keming Lu and Kai Dang and Yang Fan and Yichang Zhang and An Yang and Rui Men and Fei Huang and Bo Zheng and Yibo Miao and Shanghaoran Quan and Yunlong Feng and Xingzhang Ren and Xuancheng Ren and Jingren Zhou and Junyang Lin},
      year={2024},
      eprint={2409.12186},
      archivePrefix={arXiv},
      primaryClass={cs.CL},
      url={https://arxiv.org/abs/2409.12186}, 
}

@misc{pan2024codevbenchllmsunderstanddevelopercentric,
      title={Codev-Bench: How Do LLMs Understand Developer-Centric Code Completion?}, 
      author={Zhenyu Pan and Rongyu Cao and Yongchang Cao and Yingwei Ma and Binhua Li and Fei Huang and Han Liu and Yongbin Li},
      year={2024},
      eprint={2410.01353},
      archivePrefix={arXiv},
      primaryClass={cs.SE},
      url={https://arxiv.org/abs/2410.01353}, 
}

@misc{openai2024swebench,
  author = {{OpenAI}},
  title = {Introducing {SWE-Bench} Verified},
  howpublished = {\url{https://openai.com/index/introducing-swe-bench-verified/}},
  year = {2024},
  note = {Accessed: August 13, 2024}
}

@misc{chen2021evaluatinglargelanguagemodels,
      title={Evaluating Large Language Models Trained on Code}, 
      author={Mark Chen and Jerry Tworek and Heewoo Jun and Qiming Yuan and Henrique Ponde de Oliveira Pinto and Jared Kaplan and Harri Edwards and Yuri Burda and Nicholas Joseph and Greg Brockman and Alex Ray and Raul Puri and Gretchen Krueger and Michael Petrov and Heidy Khlaaf and Girish Sastry and Pamela Mishkin and Brooke Chan and Scott Gray and Nick Ryder and Mikhail Pavlov and Alethea Power and Lukasz Kaiser and Mohammad Bavarian and Clemens Winter and Philippe Tillet and Felipe Petroski Such and Dave Cummings and Matthias Plappert and Fotios Chantzis and Elizabeth Barnes and Ariel Herbert-Voss and William Hebgen Guss and Alex Nichol and Alex Paino and Nikolas Tezak and Jie Tang and Igor Babuschkin and Suchir Balaji and Shantanu Jain and William Saunders and Christopher Hesse and Andrew N. Carr and Jan Leike and Josh Achiam and Vedant Misra and Evan Morikawa and Alec Radford and Matthew Knight and Miles Brundage and Mira Murati and Katie Mayer and Peter Welinder and Bob McGrew and Dario Amodei and Sam McCandlish and Ilya Sutskever and Wojciech Zaremba},
      year={2021},
      eprint={2107.03374},
      archivePrefix={arXiv},
      primaryClass={cs.LG},
      url={https://arxiv.org/abs/2107.03374}, 
}

@article{cheng2024fullstack,
  title={FullStack Bench: Evaluating LLMs as Full Stack Coders},
  author={Cheng, Yao and Chen, Jianfeng and Chen, Jie and Chen, Li and Chen, Liyu and Chen, Wentao and Chen, Zhengyu and Geng, Shijie and Li, Aoyan and Li, Bo and others},
  journal={arXiv preprint arXiv:2412.00535},
  year={2024}
}

@article{jain2024livecodebench,
  title={Livecodebench: Holistic and contamination free evaluation of large language models for code},
  author={Jain, Naman and Han, King and Gu, Alex and Li, Wen-Ding and Yan, Fanjia and Zhang, Tianjun and Wang, Sida and Solar-Lezama, Armando and Sen, Koushik and Stoica, Ion},
  journal={arXiv preprint arXiv:2403.07974},
  year={2024}
}

@inproceedings{Mathews_2024, series={ASE ’24},
   title={Test-Driven Development and LLM-based Code Generation},
   url={http://dx.doi.org/10.1145/3691620.3695527},
   DOI={10.1145/3691620.3695527},
   booktitle={Proceedings of the 39th IEEE/ACM International Conference on Automated Software Engineering},
   publisher={ACM},
   author={Mathews, Noble Saji and Nagappan, Meiyappan},
   year={2024},
   month=oct, pages={1583–1594},
   collection={ASE ’24} }

@misc{ahmed2024tddbenchverifiedllmsgenerate,
      title={TDD-Bench Verified: Can LLMs Generate Tests for Issues Before They Get Resolved?}, 
      author={Toufique Ahmed and Martin Hirzel and Rangeet Pan and Avraham Shinnar and Saurabh Sinha},
      year={2024},
      eprint={2412.02883},
      archivePrefix={arXiv},
      primaryClass={cs.SE},
      url={https://arxiv.org/abs/2412.02883}, 
}

@misc{anthropic2024claude3,
  author = {Anthropic},
  title = {The Claude 3 Model Family: Opus, Sonnet, Haiku},
  year = {2024},
  url = {https://www-cdn.anthropic.com/de8ba9b01c9ab7cbabf5c33b80b7bbc618857627/Model_Card_Claude_3.pdf},
  note = {Accessed: 2024-05-10}
}

@misc{deepseekai2024deepseekcoderv2breakingbarrierclosedsource,
      title={DeepSeek-Coder-V2: Breaking the Barrier of Closed-Source Models in Code Intelligence}, 
      author={DeepSeek-AI},
      year={2024},
      eprint={2406.11931},
      archivePrefix={arXiv},
      primaryClass={cs.SE},
      url={https://arxiv.org/abs/2406.11931}, 
}

@misc{austin2021programsynthesislargelanguage,
      title={Program Synthesis with Large Language Models}, 
      author={Jacob Austin and Augustus Odena and Maxwell Nye and Maarten Bosma and Henryk Michalewski and David Dohan and Ellen Jiang and Carrie Cai and Michael Terry and Quoc Le and Charles Sutton},
      year={2021},
      eprint={2108.07732},
      archivePrefix={arXiv},
      primaryClass={cs.PL},
      url={https://arxiv.org/abs/2108.07732}, 
}

@misc{openai2024helloGPT4o,
  author = {OpenAI},
  title = {Hello GPT-4o},
  year = {2024},
  howpublished = {\url{https://openai.com/index/hello-gpt-4o/}},
  note = {Accessed: 2024-05-10}
}

@misc{seed16,
  author = {Seed},
  title = {Seed1.6 Tech Introduction},
  year = {2025},
  note = {Accessed: 2025-06-25},
  howpublished = {\url{https://seed.bytedance.com/en/seed1_6}}
}

@misc{anthropic2025claude37,
  author = {Anthropic},
  title = {Claude 3.7 Sonnet and Claude Code},
  year = {2025},
  note = {Accessed: 2025-02-25},
  howpublished = {\url{https://www.anthropic.com/news/claude-3-7-sonnet}}
}

@misc{anthropic2024claude35,
  author = {Anthropic},
  title = {Introducing Claude 3.5 Sonnet},
  year = {2024},
  note = {Accessed: 2024-06-21},
  howpublished = {\url{https://www.anthropic.com/news/claude-3-5-son-net}}
}

@article{pearson1896regression,
  author    = {Pearson, K.},
  title     = {VII. Mathematical Contributions to the Theory of Evolution.-III. Regression, Heredity, and Panmixia},
  journal   = {Philosophical Transactions of the Royal Society A},
  volume    = {187},
  pages     = {253--318},
  year      = {1896},
  doi       = {10.1098/rsta.1896.0007},
  url       = {https://doi.org/10.1098/rsta.1896.0007}
}

@misc{meta2024introducingllama31,
  author = {Meta},
  title = {Introducing Llama 3.1: Our most capable models to date},
  year = {2024},
  note = {Accessed: 2024-07-23},
  howpublished = {\url{https://ai.meta.com/blog/meta-llama-3-1/}}
}

@misc{qwen2025qwen3,
  author = {Qwen},
  title = {Qwen3: Think Deeper, Act Faster},
  year = {2025},
  note = {Accessed: 2025-04-29},
  howpublished = {\url{https://qwenlm.github.io/blog/qwen3/}}
}

@misc{qwencoder,
  author = {Qwen},
  title = {Qwen3-Coder: Agentic Coding in the World},
  year = {2025},
  note = {Accessed: 2025-07-22},
  howpublished = {\url{https://qwenlm.github.io/blog/qwen3-coder/}}
}

@article{Fasy_2014,
   title={Confidence sets for persistence diagrams},
   volume={42},
   ISSN={0090-5364},
   url={http://dx.doi.org/10.1214/14-AOS1252},
   DOI={10.1214/14-aos1252},
   number={6},
   journal={The Annals of Statistics},
   publisher={Institute of Mathematical Statistics},
   author={Fasy, Brittany Terese and Lecci, Fabrizio and Rinaldo, Alessandro and Wasserman, Larry and Balakrishnan, Sivaraman and Singh, Aarti},
   year={2014},
   month=dec 
}

@article{liu2023your,
  title={Is your code generated by chatgpt really correct? rigorous evaluation of large language models for code generation},
  author={Liu, Jiawei and Xia, Chunqiu Steven and Wang, Yuyao and Zhang, Lingming},
  journal={Advances in Neural Information Processing Systems},
  volume={36},
  pages={21558--21572},
  year={2023}
}

@article{spearman1961proof,
  title={The proof and measurement of association between two things.},
  author={Spearman, Charles},
  year={1961},
  publisher={Appleton-Century-Crofts}
}

@misc{huang2025benchmarkingllmsunittest,
      title={Benchmarking LLMs for Unit Test Generation from Real-World Functions}, 
      author={Dong Huang and Jie M. Zhang and Mark Harman and Qianru Zhang and Mingzhe Du and See-Kiong Ng},
      year={2025},
      eprint={2508.00408},
      archivePrefix={arXiv},
      primaryClass={cs.SE},
      url={https://arxiv.org/abs/2508.00408}, 
}

@article{xu2025SWECompass,
  title={SWE-Compass: Towards Unified Evaluation of Agentic Coding Abilities for Large Language Models},
  author={Xu, Jingxuan and Deng, Ken and Li, Weihao and Yu, Songwei etc},
  journal={arXiv preprint arXiv:2511.05459},
  year={2025}
}

@misc{claude45opus,
  author = {Anthropic},
  title = {Introducing Claude Opus 4.5},
  year = {2025},
  note = {Accessed: 2025-11-25},
  howpublished = {\url{https://www.anthropic.com/news/claude-opus-4-5}}
}

@misc{gpt52,
  author = {OpenAI},
  title = {Introducing GPT5.2},
  year = {2025},
  note = {Accessed: 2025-12-11},
  howpublished = {\url{https://openai.com/index/introducing-gpt-5-2/}}
}

@misc{gemini3,
  author = {Google DeepMind},
  title = {Gemini 3 Pro: Best for complex tasks and bringing creative concepts to life},
  year = {2025},
  note = {Accessed: 2025-12-5},
  howpublished = {\url{https://deepmind.google/models/gemini/pro/}}
}

@misc{wu2025repomasterevalevaluatingcodecompletion,
      title={RepoMasterEval: Evaluating Code Completion via Real-World Repositories}, 
      author={Qinyun Wu and Chao Peng and Pengfei Gao and Ruida Hu and Haoyu Gan and Bo Jiang and Jinhe Tang and Zhiwen Deng and Zhanming Guan and Cuiyun Gao and Xia Liu and Ping Yang},
      year={2025},
      eprint={2408.03519},
      archivePrefix={arXiv},
      primaryClass={cs.SE},
      url={https://arxiv.org/abs/2408.03519}, 
}

@misc{niu2023crosscodebenchbenchmarkingcrosstaskgeneralization,
      title={CrossCodeBench: Benchmarking Cross-Task Generalization of Source Code Models}, 
      author={Changan Niu and Chuanyi Li and Vincent Ng and Bin Luo},
      year={2023},
      eprint={2302.04030},
      archivePrefix={arXiv},
      primaryClass={cs.SE},
      url={https://arxiv.org/abs/2302.04030}, 
}

@misc{hai2025impactscontextsrepositorylevelcode,
      title={On the Impacts of Contexts on Repository-Level Code Generation}, 
      author={Nam Le Hai and Dung Manh Nguyen and Nghi D. Q. Bui},
      year={2025},
      eprint={2406.11927},
      archivePrefix={arXiv},
      primaryClass={cs.SE},
      url={https://arxiv.org/abs/2406.11927}, 
}

@misc{yang2024execrepobenchmultilevelexecutablecode,
      title={ExecRepoBench: Multi-level Executable Code Completion Evaluation}, 
      author={Jian Yang and Jiajun Zhang and Jiaxi Yang and Ke Jin and Lei Zhang and Qiyao Peng and Ken Deng and Yibo Miao and Tianyu Liu and Zeyu Cui and Binyuan Hui and Junyang Lin},
      year={2024},
      eprint={2412.11990},
      archivePrefix={arXiv},
      primaryClass={cs.CL},
      url={https://arxiv.org/abs/2412.11990}, 
}

@misc{li2024promptinglargelanguagemodels,
      title={Prompting Large Language Models to Tackle the Full Software Development Lifecycle: A Case Study}, 
      author={Bowen Li and Wenhan Wu and Ziwei Tang and Lin Shi and John Yang and Jinyang Li and Shunyu Yao and Chen Qian and Binyuan Hui and Qicheng Zhang and Zhiyin Yu and He Du and Ping Yang and Dahua Lin and Chao Peng and Kai Chen},
      year={2024},
      eprint={2403.08604},
      archivePrefix={arXiv},
      primaryClass={cs.CL},
      url={https://arxiv.org/abs/2403.08604}, 
}

@misc{wang2025repotransbenchrealworldmultilingualbenchmark,
      title={RepoTransBench: A Real-World Multilingual Benchmark for Repository-Level Code Translation}, 
      author={Yanli Wang and Yanlin Wang and Suiquan Wang and Daya Guo and Jiachi Chen and John Grundy and Xilin Liu and Yuchi Ma and Mingzhi Mao and Hongyu Zhang and Zibin Zheng},
      year={2025},
      eprint={2412.17744},
      archivePrefix={arXiv},
      primaryClass={cs.SE},
      url={https://arxiv.org/abs/2412.17744}, 
}

@misc{liu2025migrationbenchrepositorylevelcodemigration,
      title={MigrationBench: Repository-Level Code Migration Benchmark from Java 8}, 
      author={Linbo Liu and Xinle Liu and Qiang Zhou and Lin Chen and Yihan Liu and Hoan Nguyen and Behrooz Omidvar-Tehrani and Xi Shen and Jun Huan and Omer Tripp and Anoop Deoras},
      year={2025},
      eprint={2505.09569},
      archivePrefix={arXiv},
      primaryClass={cs.SE},
      url={https://arxiv.org/abs/2505.09569}, 
}

@misc{cassano2022multiplescalableextensibleapproach,
      title={MultiPL-E: A Scalable and Extensible Approach to Benchmarking Neural Code Generation}, 
      author={Federico Cassano and John Gouwar and Daniel Nguyen and Sydney Nguyen and Luna Phipps-Costin and Donald Pinckney and Ming-Ho Yee and Yangtian Zi and Carolyn Jane Anderson and Molly Q Feldman and Arjun Guha and Michael Greenberg and Abhinav Jangda},
      year={2022},
      eprint={2208.08227},
      archivePrefix={arXiv},
      primaryClass={cs.LG},
      url={https://arxiv.org/abs/2208.08227}, 
}

@misc{ibrahimzada2025challengingbugpredictionrepair,
      title={Challenging Bug Prediction and Repair Models with Synthetic Bugs}, 
      author={Ali Reza Ibrahimzada and Yang Chen and Ryan Rong and Reyhaneh Jabbarvand},
      year={2025},
      eprint={2310.02407},
      archivePrefix={arXiv},
      primaryClass={cs.SE},
      url={https://arxiv.org/abs/2310.02407}, 
}

@misc{liu2023repobenchbenchmarkingrepositorylevelcode,
      title={RepoBench: Benchmarking Repository-Level Code Auto-Completion Systems}, 
      author={Tianyang Liu and Canwen Xu and Julian McAuley},
      year={2023},
      eprint={2306.03091},
      archivePrefix={arXiv},
      primaryClass={cs.CL},
      url={https://arxiv.org/abs/2306.03091}, 
}

@misc{li2025feabenchbenchmarkevaluatingrepositorylevel,
      title={FEA-Bench: A Benchmark for Evaluating Repository-Level Code Generation for Feature Implementation}, 
      author={Wei Li and Xin Zhang and Zhongxin Guo and Shaoguang Mao and Wen Luo and Guangyue Peng and Yangyu Huang and Houfeng Wang and Scarlett Li},
      year={2025},
      eprint={2503.06680},
      archivePrefix={arXiv},
      primaryClass={cs.SE},
      url={https://arxiv.org/abs/2503.06680}, 
}

@misc{kwon2023efficientmemorymanagementlarge,
      title={Efficient Memory Management for Large Language Model Serving with PagedAttention}, 
      author={Woosuk Kwon and Zhuohan Li and Siyuan Zhuang and Ying Sheng and Lianmin Zheng and Cody Hao Yu and Joseph E. Gonzalez and Hao Zhang and Ion Stoica},
      year={2023},
      eprint={2309.06180},
      archivePrefix={arXiv},
      primaryClass={cs.LG},
      url={https://arxiv.org/abs/2309.06180}, 
}

@misc{deepseekv32,
      title={DeepSeek-V3.2: Pushing the Frontier of Open Large Language Models}, 
      author={DeepSeek-AI and Aixin Liu and Aoxue Mei and Bangcai Lin and Bing Xue and Bingxuan Wang and Bingzheng Xu and Bochao Wu and Bowei Zhang and Chaofan Lin and Chen Dong and Chengda Lu and Chenggang Zhao and Chengqi Deng and Chenhao Xu and Chong Ruan and Damai Dai and Daya Guo and Dejian Yang and Deli Chen and Erhang Li and Fangqi Zhou and Fangyun Lin and Fucong Dai and Guangbo Hao and Guanting Chen and Guowei Li and H. Zhang and Hanwei Xu and Hao Li and Haofen Liang and Haoran Wei and Haowei Zhang and Haowen Luo and Haozhe Ji and Honghui Ding and Hongxuan Tang and Huanqi Cao and Huazuo Gao and Hui Qu and Hui Zeng and Jialiang Huang and Jiashi Li and Jiaxin Xu and Jiewen Hu and Jingchang Chen and Jingting Xiang and Jingyang Yuan and Jingyuan Cheng and Jinhua Zhu and Jun Ran and Junguang Jiang and Junjie Qiu and Junlong Li and Junxiao Song and Kai Dong and Kaige Gao and Kang Guan and Kexin Huang and Kexing Zhou and Kezhao Huang and Kuai Yu and Lean Wang and Lecong Zhang and Lei Wang and Liang Zhao and Liangsheng Yin and Lihua Guo and Lingxiao Luo and Linwang Ma and Litong Wang and Liyue Zhang and M. S. Di and M. Y Xu and Mingchuan Zhang and Minghua Zhang and Minghui Tang and Mingxu Zhou and Panpan Huang and Peixin Cong and Peiyi Wang and Qiancheng Wang and Qihao Zhu and Qingyang Li and Qinyu Chen and Qiushi Du and Ruiling Xu and Ruiqi Ge and Ruisong Zhang and Ruizhe Pan and Runji Wang and Runqiu Yin and Runxin Xu and Ruomeng Shen and Ruoyu Zhang and S. H. Liu and Shanghao Lu and Shangyan Zhou and Shanhuang Chen and Shaofei Cai and Shaoyuan Chen and Shengding Hu and Shengyu Liu and Shiqiang Hu and Shirong Ma and Shiyu Wang and Shuiping Yu and Shunfeng Zhou and Shuting Pan and Songyang Zhou and Tao Ni and Tao Yun and Tian Pei and Tian Ye and Tianyuan Yue and Wangding Zeng and Wen Liu and Wenfeng Liang and Wenjie Pang and Wenjing Luo and Wenjun Gao and Wentao Zhang and Xi Gao and Xiangwen Wang and Xiao Bi and Xiaodong Liu and Xiaohan Wang and Xiaokang Chen and Xiaokang Zhang and Xiaotao Nie and Xin Cheng and Xin Liu and Xin Xie and Xingchao Liu and Xingkai Yu and Xingyou Li and Xinyu Yang and Xinyuan Li and Xu Chen and Xuecheng Su and Xuehai Pan and Xuheng Lin and Xuwei Fu and Y. Q. Wang and Yang Zhang and Yanhong Xu and Yanru Ma and Yao Li and Yao Li and Yao Zhao and Yaofeng Sun and Yaohui Wang and Yi Qian and Yi Yu and Yichao Zhang and Yifan Ding and Yifan Shi and Yiliang Xiong and Ying He and Ying Zhou and Yinmin Zhong and Yishi Piao and Yisong Wang and Yixiao Chen and Yixuan Tan and Yixuan Wei and Yiyang Ma and Yiyuan Liu and Yonglun Yang and Yongqiang Guo and Yongtong Wu and Yu Wu and Yuan Cheng and Yuan Ou and Yuanfan Xu and Yuduan Wang and Yue Gong and Yuhan Wu and Yuheng Zou and Yukun Li and Yunfan Xiong and Yuxiang Luo and Yuxiang You and Yuxuan Liu and Yuyang Zhou and Z. F. Wu and Z. Z. Ren and Zehua Zhao and Zehui Ren and Zhangli Sha and Zhe Fu and Zhean Xu and Zhenda Xie and Zhengyan Zhang and Zhewen Hao and Zhibin Gou and Zhicheng Ma and Zhigang Yan and Zhihong Shao and Zhixian Huang and Zhiyu Wu and Zhuoshu Li and Zhuping Zhang and Zian Xu and Zihao Wang and Zihui Gu and Zijia Zhu and Zilin Li and Zipeng Zhang and Ziwei Xie and Ziyi Gao and Zizheng Pan and Zongqing Yao and Bei Feng and Hui Li and J. L. Cai and Jiaqi Ni and Lei Xu and Meng Li and Ning Tian and R. J. Chen and R. L. Jin and S. S. Li and Shuang Zhou and Tianyu Sun and X. Q. Li and Xiangyue Jin and Xiaojin Shen and Xiaosha Chen and Xinnan Song and Xinyi Zhou and Y. X. Zhu and Yanping Huang and Yaohui Li and Yi Zheng and Yuchen Zhu and Yunxian Ma and Zhen Huang and Zhipeng Xu and Zhongyu Zhang and Dongjie Ji and Jian Liang and Jianzhong Guo and Jin Chen and Leyi Xia and Miaojun Wang and Mingming Li and Peng Zhang and Ruyi Chen and Shangmian Sun and Shaoqing Wu and Shengfeng Ye and T. Wang and W. L. Xiao and Wei An and Xianzu Wang and Xiaowen Sun and Xiaoxiang Wang and Ying Tang and Yukun Zha and Zekai Zhang and Zhe Ju and Zhen Zhang and Zihua Qu},
      year={2025},
      eprint={2512.02556},
      archivePrefix={arXiv},
      primaryClass={cs.CL},
      url={https://arxiv.org/abs/2512.02556}, 
}

@misc{glm46,
  author = {Z.ai},
  title = {GLM-4.6: Advanced Agentic, Reasoning and Coding Capabilities},
  year = {2025},
  note = {Accessed: 2025-09-30},
  howpublished = {\url{https://z.ai/blog/glm-4.6}}
}

@misc{kimik2,
      title={Kimi K2: Open Agentic Intelligence}, 
      author={Kimi and Yifan Bai and Yiping Bao and Guanduo Chen and Jiahao Chen and Ningxin Chen and Ruijue Chen and Yanru Chen and Yuankun Chen and Yutian Chen and Zhuofu Chen and Jialei Cui and Hao Ding and Mengnan Dong and Angang Du and Chenzhuang Du and Dikang Du and Yulun Du and Yu Fan and Yichen Feng and Kelin Fu and Bofei Gao and Hongcheng Gao and Peizhong Gao and Tong Gao and Xinran Gu and Longyu Guan and Haiqing Guo and Jianhang Guo and Hao Hu and Xiaoru Hao and Tianhong He and Weiran He and Wenyang He and Chao Hong and Yangyang Hu and Zhenxing Hu and Weixiao Huang and Zhiqi Huang and Zihao Huang and Tao Jiang and Zhejun Jiang and Xinyi Jin and Yongsheng Kang and Guokun Lai and Cheng Li and Fang Li and Haoyang Li and Ming Li and Wentao Li and Yanhao Li and Yiwei Li and Zhaowei Li and Zheming Li and Hongzhan Lin and Xiaohan Lin and Zongyu Lin and Chengyin Liu and Chenyu Liu and Hongzhang Liu and Jingyuan Liu and Junqi Liu and Liang Liu and Shaowei Liu and T. Y. Liu and Tianwei Liu and Weizhou Liu and Yangyang Liu and Yibo Liu and Yiping Liu and Yue Liu and Zhengying Liu and Enzhe Lu and Lijun Lu and Shengling Ma and Xinyu Ma and Yingwei Ma and Shaoguang Mao and Jie Mei and Xin Men and Yibo Miao and Siyuan Pan and Yebo Peng and Ruoyu Qin and Bowen Qu and Zeyu Shang and Lidong Shi and Shengyuan Shi and Feifan Song and Jianlin Su and Zhengyuan Su and Xinjie Sun and Flood Sung and Heyi Tang and Jiawen Tao and Qifeng Teng and Chensi Wang and Dinglu Wang and Feng Wang and Haiming Wang and Jianzhou Wang and Jiaxing Wang and Jinhong Wang and Shengjie Wang and Shuyi Wang and Yao Wang and Yejie Wang and Yiqin Wang and Yuxin Wang and Yuzhi Wang and Zhaoji Wang and Zhengtao Wang and Zhexu Wang and Chu Wei and Qianqian Wei and Wenhao Wu and Xingzhe Wu and Yuxin Wu and Chenjun Xiao and Xiaotong Xie and Weimin Xiong and Boyu Xu and Jing Xu and Jinjing Xu and L. H. Xu and Lin Xu and Suting Xu and Weixin Xu and Xinran Xu and Yangchuan Xu and Ziyao Xu and Junjie Yan and Yuzi Yan and Xiaofei Yang and Ying Yang and Zhen Yang and Zhilin Yang and Zonghan Yang and Haotian Yao and Xingcheng Yao and Wenjie Ye and Zhuorui Ye and Bohong Yin and Longhui Yu and Enming Yuan and Hongbang Yuan and Mengjie Yuan and Haobing Zhan and Dehao Zhang and Hao Zhang and Wanlu Zhang and Xiaobin Zhang and Yangkun Zhang and Yizhi Zhang and Yongting Zhang and Yu Zhang and Yutao Zhang and Yutong Zhang and Zheng Zhang and Haotian Zhao and Yikai Zhao and Huabin Zheng and Shaojie Zheng and Jianren Zhou and Xinyu Zhou and Zaida Zhou and Zhen Zhu and Weiyu Zhuang and Xinxing Zu},
      year={2025},
      eprint={2507.20534},
      archivePrefix={arXiv},
      primaryClass={cs.LG},
      url={https://arxiv.org/abs/2507.20534}, 
}

@misc{minimaxm2,
  author = {MINIMAX},
  title = {MiniMax M2 \& Agent: Ingenious in Simplicity},
  year = {2025},
  note = {Accessed: 2025-10-27},
  howpublished = {\url{https://www.minimax.io/news/minimax-m2}}
}

\newpage

\appendix
\section{Source Repository Details}\label{app:repo}

Table~\ref{tab:repo_info} summarizes the metadata and statistical characteristics of the 12 selected repositories. To ensure the benchmark's representativeness and quality, we curated these projects from PyPI\footnote{https://pypi.org/} based on strict inclusion criteria. As shown in the table, the selected repositories exhibit significant diversity in domain, test framework (e.g., \texttt{unittest} vs \texttt{pytest}), and project structure. 
Despite this heterogeneity, \textsc{CorePipe} successfully parses and processes all repositories, demonstrating robust adaptability to varied engineering conventions.

\begin{table*}[htb]
\caption{Repository Information.}
\label{tab:repo_info}
\centering
\resizebox{\textwidth}{!}{%
\begin{tabular}{lcccccccc}
\toprule
\textbf{Repo} & \textbf{Created Time} & \textbf{Latest Version} & \textbf{Latest Release Time} & \textbf{Github Link} & \textbf{Total Code Lines} & \textbf{Python Files} & \textbf{Test Files} & \textbf{Test Coverage (\%)} \\
\midrule
transformers & 2019/9/26 & 4.51.3 & 2025/4/14 & \href{https://github.com/huggingface/transformers}{/huggingface/transformers} & 971,687 & 1,756 & 712 & 40.55 \\
langchain & 2022/10/25 & 0.3.25 & 2025/5/3 & \href{https://github.com/langchain-ai/langchain}{/langchain-ai/langchain} & 68,790 & 1,329 & 265 & 19.94 \\
datachain & 2024/6/27 & 0.16.4 & 2025/5/1 & \href{https://github.com/iterative/datachain/tree/main}{/iterative/datachain/tree/main} & 26,777 & 137 & 57 & 41.61 \\
open-iris & 2023/12/14 & 1.5.0 & 2025/4/22 & \href{https://github.com/worldcoin/open-iris}{/worldcoin/open-iris} & 8,072 & 76 & 64 & 84.21 \\
UniRef & 2023/12/26 & 0.6 & 2023/12/26 & \href{https://github.com/FoundationVision/UniRef}{/FoundationVision/UniRef} & 36,127 & 152 & 50 & 32.89 \\
haystack & 2023/11/25 & 2.13.1 & 2025/4/24 & \href{https://github.com/deepset-ai/haystack}{/deepset-ai/haystack} & 33,905 & 211 & 150 & 71.09 \\
d3rlpy & 2020/7/31 & 2.8.1 & 2025/3/2 & \href{https://github.com/takuseno/d3rlpy}{/takuseno/d3rlpy} & 23,984 & 125 & 45 & 36.00 \\
inference & 2023/8/16 & 0.48.3 & 2025/5/6 & \href{https://github.com/roboflow/inference}{/roboflow/inference} & 83,164 & 640 & 118 & 18.44 \\
rdt & 2018/8/23 & 1.16.0 & 2025/4/11 & \href{https://github.com/sdv-dev/RDT}{/sdv-dev/RDT} & 7,265 & 31 & 16 & 51.61 \\
cloudnetpy & 2019/9/13 & 1.75.0 & 2025/5/2 & \href{https://github.com/actris-cloudnet/cloudnetpy}{/actris-cloudnet/cloudnetpy} & 23,025 & 116 & 49 & 42.24 \\
skfolio & 2023/12/15 & 0.9.0 & 2025/4/5 & \href{https://github.com/skfolio/skfolio}{/skfolio/skfolio} & 29,865 & 113 & 71 & 62.83 \\
finam & 2023/2/3 & 1.0.1 & 2025/4/23 & \href{https://github.com/finam-ufz/finam}{/finam-ufz/finam} & 12,592 & 46 & 30 & 65.22 \\
\bottomrule
\end{tabular}
}
\end{table*}

\section{Robustness of Benchmarking Results to Generator Selection}\label{app:model-selection}

\begin{figure*}[t]
    \centering
    \includegraphics[width=\linewidth]{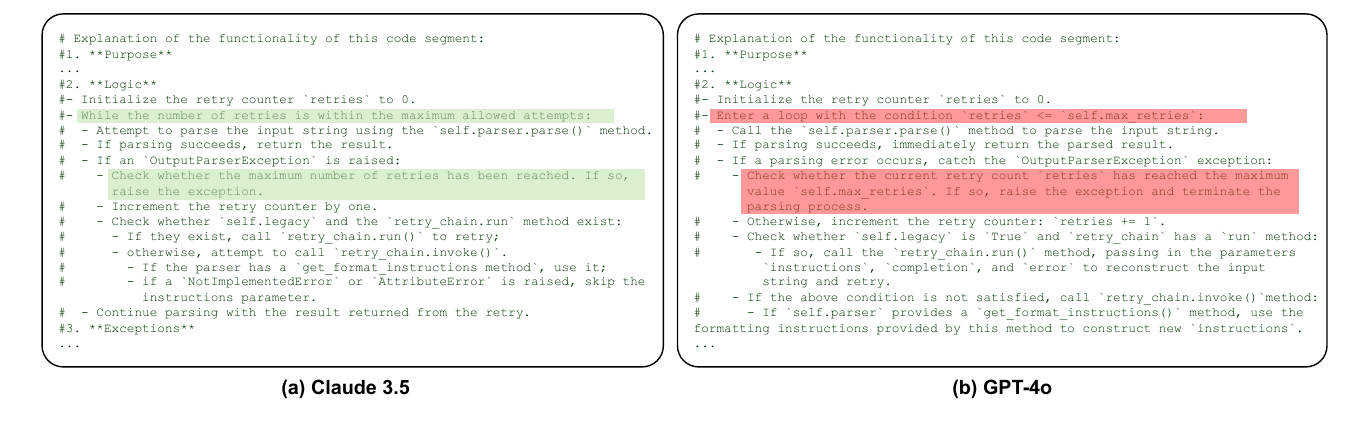}
    \caption{\textbf{Qualitative comparison of generated descriptions.} GPT-4o (b) captures more granular logic details (highlighted in red) compared to the more concise Claude 3.5 (a). {Despite these significant stylistic divergences, our robustness analysis confirms that model rankings remain stable.}}
    \label{fig:compareBackbone}
\end{figure*}

We utilize GPT-4o as the primary backbone to generate structured functional descriptions. To verify that our evaluation results are not biased by this choice, we conduct a comprehensive sensitivity analysis using four distinct generator backbones: GPT-4o, Claude-3.5-Sonnet, Qwen-Plus-Latest, and Doubao-Pro-4k. As visualized in Figure~\ref{fig:compareBackbone}, different backbone models exhibit distinct stylistic characteristics: GPT-4o tends to produce richer, more granular details (e.g., explicit loop conditions), whereas Claude 3.5 favors conciseness.

\begin{figure}[htbp]
    \centering
    \includegraphics[width=\linewidth]{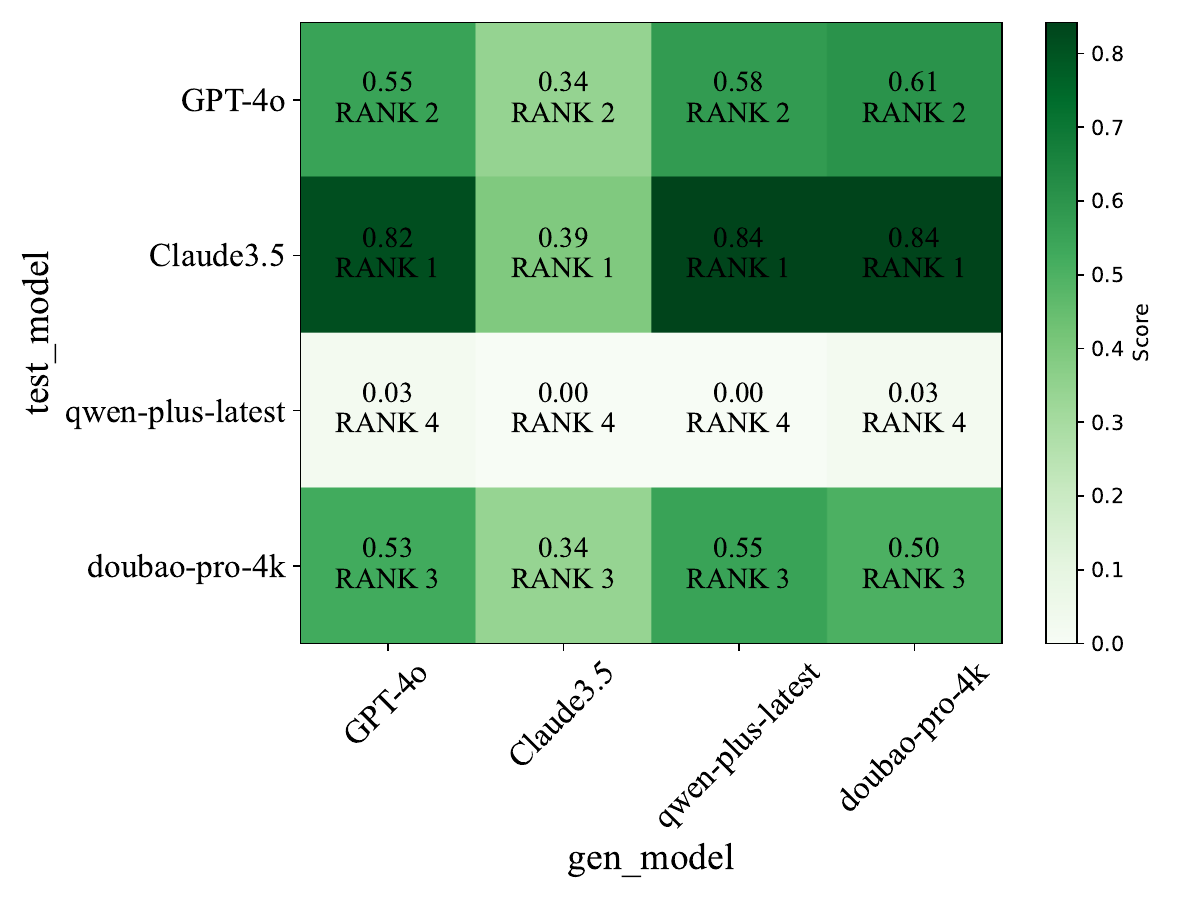}
    \caption{\textbf{Generator Sensitivity Analysis.} Performance heatmap of evaluated models across descriptions generated by different backbones. While absolute pass rates fluctuate due to prompt style, the relative ranking of models remains highly consistent, confirming the benchmark's robustness to generator choice.} 
    \label{fig:backboneComp}
\end{figure}

Despite this variation in input prompt density, our quantitative analysis demonstrates that the benchmark's discriminative power remains invariant. As illustrated in Figure~\ref{fig:backboneComp}, we evaluate four representative test models across tasks generated by each backbone. 
While absolute pass rates fluctuate slightly due to varying description styles (e.g., Claude-3.5 tends to be more concise), the relative ranking of the test models remains strictly invariant.
Specifically, regardless of which model generated the task, the performance order is consistently: Claude-3.5 $>$ GPT-4o $>$ Doubao-Pro $>$ Qwen-Plus-Latest.
This yields a perfect Spearman Rank Correlation Coefficient of ${\rho = 1.0}$ across all generator configurations.
This empirical evidence strongly confirms that \textsc{CoreCodeBench} provides a stable, generator-agnostic assessment of coding proficiency.

Furthermore, we evaluate the performance of different models on development-type tasks using descriptions generated by different backbones. As illustrated in Figure~\ref{fig:backboneComp} the absolute scores of the models fluctuate due to differences in the description styles. However, the relative ranking of the models remains largely consistent and is not affected by the choice of the backbone model.

\section{Information Gain (IG) Filtering Protocol}\label{app:igeval}

To ensure that the generated natural language descriptions in development tasks provide effective guidance rather than noise, we introduce an Information Gain (IG) Score. 
We define the score for a given problem as:
\begin{equation}
\text{IG} = \text{AC Rate}_{\text{exp}} - \text{AC Rate}_{\text{no-exp}}
\end{equation}
where $\text{AC Rate}_{\text{exp}}$ is the average pass rate of baseline models (GPT-4o, Claude-3.5, Doubao-pro-4k, Qwen-plus-latest) when provided with the generated description, and $\text{AC Rate}_{\text{no-exp}}$ is the pass rate when provided only with the masked code context (i.e., infilling mode).

\paragraph{Filtering Criteria.}
We retain a problem if it satisfies one of the following conditions:
\begin{enumerate}[leftmargin=10pt]
    \item $\text{IG} > 0$: The description positively aids the model, indicating it contains valid functional specifications.
    \item $\text{AC Rate}_{\text{exp}} = \text{AC Rate}_{\text{no-exp}} = 0$: The problem is challenging for all baselines regardless of the description. We retain these as "Hard" instances, provided there is no evidence that the description is misleading (i.e., we filter out cases where $\text{AC Rate}_{\text{no-exp}} > \text{AC Rate}_{\text{exp}}$).
\end{enumerate}
After applying this filter, 48.56\% of the generated candidates are retained.

\paragraph{Ablation Study.}
As shown in Table~\ref{tab:igmodel}, we evaluate models on the discarded subset (IG $\leq 0$). The results show compressed performance gaps and random fluctuations, confirming that low-IG questions fail to effectively discriminate model capabilities.

\begin{table}[h]
    \caption{Model Performance on Low-IG Problems.}
    \label{tab:igmodel}
    \centering
    \small
    \begin{tabular}{lccc}
        \toprule
        \textbf{Model}   &\textbf{AC Rate} & \textbf{AC@1} \\ \midrule
        {Gemini-2.5-Pro} & 98.84 & 92.52 \\
        {GPT-5} & 98.23 & 92.53\\
        {Doubao-Seed-1.6} & 85.11 & 71.28\\
        {qwen-plus-latest} & 92.90 & 82.11 \\
        \bottomrule
    \end{tabular}
\end{table}

\section{Qualitative Analysis of Bug Realism}\label{app:bugfix-casestudy}

To ensure the generated bugs reflect real-world development scenarios rather than artificial noise, we validate our cascaded logic-implementation synthesis approach.

\paragraph{Rationale for Cascaded Synthesis.}
Directly prompting advanced models (e.g., GPT-4o) to generate buggy code often results in \textit{resistance to failure} due to RLHF alignment, or yields trivial syntax errors when forced. 
By decoupling logical fallacy design (assigned to advanced models for complex error planning, including gpt-4o, claude 3.5, etc.) from implementation (assigned to smaller models, such as QwenCoder 30B, DeepSeek 16B), we simulate a realistic scenario: a developer conceptualizing a solution but encountering implementation discrepancies. 
This two-stage approach leverages the reasoning depth of large models to ensure logical complexity, while utilizing the stochasticity of smaller models to introduce natural implementation noise (e.g., variable shadowing, boundary oversights), thereby maintaining the semantic structure of the code.

\paragraph{Bug Diversity and Characteristics.}
Systematic analysis reveals a diverse distribution of error types mirroring realistic developer mistakes:
\begin{itemize}[leftmargin=10pt]
    \item \textbf{Boundary Value Errors}: Off-by-one errors in loops or conditional checks (e.g., \texttt{<} vs \texttt{<=}).
    \item \textbf{Control Flow Logic}: Incorrect nesting, premature \texttt{return}, or missing termination conditions.
    \item \textbf{Variable Misuse}: Variable shadowing, state update errors, or incorrect variable references.
    \item \textbf{Exception Handling}: Omission of edge case checks (e.g., \texttt{None} types).
    \item \textbf{Algorithm Logic Errors}: Logic inversion or incorrect operator precedence.
\end{itemize}
This diversity significantly outperforms simple syntactic perturbation baselines in terms of realism and semantic coherence.

\paragraph{Case Study Comparison.}
We compare a bug generated by a direct GPT-4o prompt versus one by our \textsc{CorePipe} framework.

\noindent\textbf{Original Correct Code:}
\begin{pycode}
size = get_size_dict(size)
shortest_edge = min(size["height"], size["width"])

output_size = get_resize_output_image_size(
    image, 
    size=shortest_edge, 
    default_to_square=False, 
    input_data_format=input_data_format
)

resized_image = resize(
    image,
    size=output_size,
    resample=resample,
    data_format=data_format,
    input_data_format=input_data_format,
    **kwargs,
)
\end{pycode}

\noindent\textbf{Baseline (GPT-4o Direct Generation):}
The model introduces a trivial operator inversion (\texttt{min} $\rightarrow$ \texttt{max}). This represents a superficial error lacking semantic depth.

\begin{pycode}
# Artificial Error: Simple Operator Flip
shortest_edge = max(size["height"], size["width"]) 
\end{pycode}

\noindent\textbf{\textsc{CorePipe} Generation (Ours):}
The cascaded approach synthesizes a compound error involving API hallucination (\texttt{size["shortest\_edge"]}), parameter misalignment (\texttt{is\_square}), and variable shadowing. These errors reflect plausible semantic misunderstandings typical of complex implementation tasks.

\begin{pycode}[breaklines=true]
# Realistic Developer Slips: API Misuse, Key Error, and Variable Shadowing
resized_image = get_resize_output_image_size(image, size["shortest_edge"], is_square=False, input_data_format=input_data_format)

# Superfluous Logic
if resized_image[0] > resized_image[1]:
    resized_image = resized_image[::-1]

resized_image = resize(image, size=resized_image, resample=resample, data_format=data_format, input_data_format=input_data_format)
\end{pycode}

Finally, consistent with our validity protocols, all generated buggy codes undergo strict unit test regression validation. Only code that passes syntax checks but fails specific functional tests is retained, ensuring the bugs are executable, non-trivial, and theoretically repairable.

\section{Generation Rules for Composite Tasks}\label{app:multi-problem-rules}

To construct composite tasks that simulate realistic engineering workflows, we employ a rigorous subgraph sampling and node assignment protocol. The generation process follows three key steps:

\paragraph{Subgraph Sampling.}
We first construct the global Function Call Tree $\mathcal{G} = (V, E)$ for each repository. 
Given the \textbf{configurable hyperparameters} for task quantity ($\nu$) and dependency depth ($d$), we randomly sample a connected subgraph $\mathcal{G}' \subseteq \mathcal{G}$ such that $|V(\mathcal{G}')| = \nu$ and the depth of $\mathcal{G}'$ does not exceed $d$.
 Crucially, we verify that $\mathcal{G}'$ remains a valid call hierarchy where the root node's execution trace can reach all leaf nodes, ensuring that \textit{no node is isolated from the testing context}.

\paragraph{Node Assignment Rules.}
Once the subgraph structure is determined, we assign specific atomic task types to each node $v \in V(\mathcal{G}')$ based on the composite task category:

\begin{itemize}[leftmargin=10pt]
    \item \textbf{Multi-Dev}: 
    All nodes are assigned as \textit{Development} tasks. To simulate dependency implementation, leaf nodes or auxiliary utilities may be converted into \textit{Empty-Function} nodes (signature only), forcing the model to implement supporting logic from scratch.
    
    \item \textbf{Multi-BugFix}: 
    All nodes are assigned as \textit{BugFix} tasks. We strictly prohibit mixing Development tasks here, adhering to the real-world principle that debugging sessions typically focus on rectifying existing logic rather than implementing new features simultaneously.
    
    \item \textbf{Multi-TDD}: 
    At least one node (typically the root or a core logic node) is assigned as a \textit{TDD} task (implementation via test constraints). Remaining nodes can be \textit{Development} or \textit{Empty-Function} tasks. This hybrid composition simulates a collaborative environment where developers must align new implementations with rigid test specifications while managing dependencies.
\end{itemize}

\paragraph{Difficult Subset.}
For the \textsc{CoreCodeBench}-\textit{Difficult} subset ($\nu=+\infty$), we enforce a minimum complexity constraint of $n \ge 3$ nodes, ensuring that the task requires reasoning over a non-trivial dependency chain.


\section{Illustrative Examples of Atomic Tasks}\label{app:datasource}

\begin{figure*}[h]
    \centering
    \includegraphics[width=\linewidth]{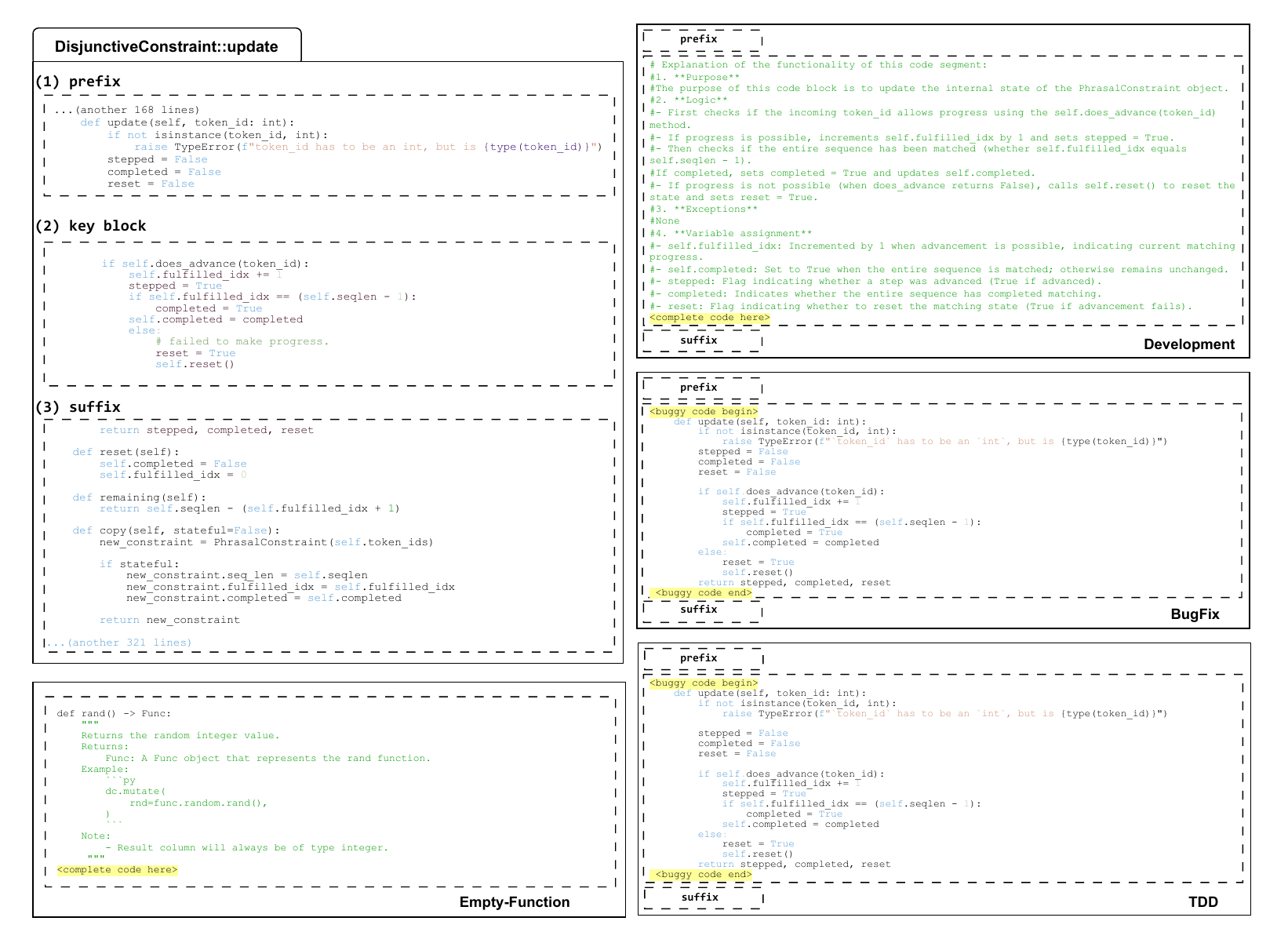}
    \caption{Illustration of atomic single-function problems.}
    \label{fig:illustration}
\end{figure*}

Figure~\ref{fig:illustration} visualizes the structure of the four atomic task types: {Development}, {BugFix}, {TDD}, and {Empty-Function}. 
As shown, all tasks are derived from the same function context but mask different components to isolate distinct cognitive demands:
\begin{itemize}[leftmargin=10pt]
    \item \textbf{Development}: Masks the function body, providing a natural language spec.
    \item \textbf{BugFix}: Provides a complete but logically flawed implementation.
    \item \textbf{TDD}: Masks the body but provides unit tests as the specification.
    \item \textbf{Empty-Function}: Specialized for composite tasks, this type strips the body to a bare signature, serving as a dependency node that must be implemented synchronously.
\end{itemize}
Composite tasks (multi-function problems) aggregate these atomic units into a unified prompt format, maintaining consistency in input/output interfaces.

\section{Prompts of Evaluation}\label{app:problemprompt}
Below, we present the evaluation prompts used for each of the six problem types.

\subsection{Prompt for Atomic Tasks}

\textbf{Single-Development}

\begin{lstlisting}[breaklines=true]
Below is a code snippet containing a placeholder `<complete code here>`. Please analyze the provided context and description of the missing code to generate the appropriate code block at `<complete code here>`.
Please output the completed code block using markdown format (```python```).
**Important**: Ensure the code block you complete maintains the same indentation as the context code, meaning you need to preserve the original code's indentation. The output must exactly match the line count and structure of the input, including preserving empty lines and comment positions.
Code snippet:
```python
{prompt}
```
Please output the completed code block using markdown format (```python```). Make sure to preserve the original indentation before and after the <complete code here> placeholder. And remember don't add the signature of the function into it.
\end{lstlisting}

\textbf{Single-BugFix}

\begin{lstlisting}[breaklines=true]
In the following code snippet, there is a buggy code section between `<buggy code begin>` and `<buggy code end>`. I've provided the corresponding unit test file and pytest error messages. Please analyze the given context and rewrite the erroneous code segment.
Please format the rewritten function block in markdown (```python```), including only the rewritten content between `<buggy code begin>` and `<buggy code end>`, without including the `<buggy code begin>` and `<buggy code end>` tags.
**Note**: Please ensure that your completed code block maintains the indentation of the original code context.
Code snippet:
```python
{new_code}
```
Unit test code:
```python
{test_code}
```
Test error log:
```
{log}
```
\end{lstlisting}

\textbf{Single-TDD}

\begin{lstlisting}[breaklines=true]
Below is a code file {file_name} containing a placeholder `<complete code here>`.Please analyze the provided file context and unit test information, and generate appropriate code at the `<complete code here>` location. Please output your completed code block in markdown format (```python```). The code block should only include the code at the `<completed code here>` location, without the surrounding context.
**Note**: Please ensure that your completed code block maintains the indentation of the surrounding code, meaning you need to preserve the original code's indentation.

Code file {file_name} to be completed:
```python
{new_code}
```
Corresponding unit test:
```python
{test_file}
```
\end{lstlisting}

\subsection{Prompts for Composite Tasks}
\paragraph{Output Format and Grading Logic for Composite Tasks}

To enable multi-function editing, we enforce a structured output format where models sequentially implement target functions wrapped in explicit ID tags (e.g., \texttt{<id> code </id>}). 
Our evaluation harness automatically \textbf{extracts} these tagged segments, \textbf{patches} them into the original repository context, and \textbf{executes} the test suite for verification.

\textbf{Multi-Development}

\begin{lstlisting}[breaklines=true]
You are a code completion agent, I would provide you with a snippet of code, and you would need to return the completed code segment. 
the code after <ralated code> is used while calling the code to be completed. 
You need to complete code blocks after <complete following code> by predicting the codes after <complete code here>, <id> label wraps the position of the code.
Your output should include the <id></id> label, followed by the completed code snippet enclosed within triple backticks ```, ensuring clarity and proper formatting.

<related code>
<id>{id}<\id>
{related code}

<complete following code>
<id>{id}<\id>
{function code}

\end{lstlisting}

\textbf{Multi-BugFix}

\begin{lstlisting}[breaklines=true]
In the following code snippet, the code between <buggy code begin> and <buggy code end> contains bugs, <id> label wraps the position of the code. Please analyze the provided context and rewrite the faulty code segment.
The code after <related code> is used while calling the code to be rewritten. 
Your output should include the <id></id> label, followed by the new code snippet enclosed within triple backticks ```, ensuring clarity and proper formatting.

<related code>
<id>{id}<\id>
{related code}

<complete following code>
<id>{id}<\id>
{function code}
\end{lstlisting}

\textbf{Multi-TDD}

\begin{lstlisting}[breaklines=true]
You are a code completion agent, I would provide you with a snippet of code, and you would need to return the completed code segment. 
The code after <ralated code> is used while calling the code to be completed. 
You need to complete code blocks after <complete following code> by predicting the codes after <complete code here>, <id> label wraps the position of the code.
Please analyze the provided file context and the unit test information of the file, and generate an appropriate code block at the position marked <complete code here>.
Your output should include the <id></id> label, followed by the completed code snippet enclosed within triple backticks ```, ensuring clarity and proper formatting.
Note: Please ensure that the code block you provide as a completion matches the indentation of the surrounding context, i.e., you need to preserve the original code's indentation.

<related code>
<id>{id}<\id>
{related code}

<complete following code>
<id>{id}<\id>
{function code}

The unit test information:
{test_codes}
\end{lstlisting}

\section{Evaluation Robustness to Prompt Variation}\label{app:promptvariation}

\subsection{Evaluation Consistency under Prompt Variations}

\begin{table*}[htb]
    \caption{\textbf{Model Performance Stability Under Prompt Rephrasing.} MAPD means the Maximum Absolute Pairwise Difference of evaluation results between different prompts. Conf. means 95\% Confidence Interval of evaluation results.}
    \label{tab:difprompt}
    \centering
    \small
    \resizebox{\textwidth}{!}{%
    \begin{tabular}{lcccccccc}
        \toprule
        \multirow{3}{*}{\textbf{Model}} & \multicolumn{4}{c}{\textbf{Single-Function Problem}} & \multicolumn{4}{c}{\textbf{Multi-Function Problem}} \\ \cmidrule(lr){2-5} \cmidrule(lr){6-9}
        & \multicolumn{2}{c}{\textbf{AC Rate}} & \multicolumn{2}{c}{\textbf{AC@1}}& \multicolumn{2}{c}{\textbf{AC Rate}} & \multicolumn{2}{c}{\textbf{AC@1}} \\ \cmidrule(lr){2-3} \cmidrule(lr){4-5} \cmidrule(lr){6-7} \cmidrule(lr){8-9}
        &\textbf{MAPD} & \textbf{Conf.} &\textbf{MAPD} & \textbf{Conf.} &\textbf{MAPD} & \textbf{Conf.} &\textbf{MAPD} & \textbf{Conf.} \\ 
        \midrule
        {Claude-3.7-Sonnet} & 0.24 &2.35& 0.27 & 2.93 & 3.49 & 4.09  & 0.31 & 3.62 \\
        {GPT-5}  & 3.97 & 2.05 & 3.99 &2.83 & 1.16 & 4.54 & 0.61 & 4.32\\
        {Llama3.1-70B}  & 0.80 & 2.41 & 0.54 & 2.54 & 0.81 & 3.43 & 0.31 & 2.77\\
        {Qwen3-Coder}  & 1.88 & 2.24 & 1.18 & 2.95 & 1.72 & 3.41 & 2.44 & 2.65\\
        \bottomrule
    \end{tabular}}
\end{table*}

To investigate the robustness of evaluation results under prompt variation, we evaluate four mainstream LLMs using three distinct rephrased prompts for each task. Table~\ref{tab:difprompt} reports the Maximum Absolute Pairwise Difference (MAPD) across these variations compared to the model's 95\% Confidence Interval in \textsc{CoreCodeBench}. For the majority of models, the performance variation (MAPD) is strictly smaller than the inherent statistical uncertainty (Confidence Interval)~\citep{Fasy_2014}. This suggests that \textsc{CoreCodeBench} provides stable assessments that are robust to surface-level prompt changes.
In the case of GPT-5, slightly larger fluctuations are observed, which we attribute to the inherent non-determinism of the model's API even at temperature 0, rather than benchmark instability. Nonetheless, these variations remain within acceptable bounds for ranking purposes.

\subsection{Consistency under Context Length Variation}

\begin{figure}[h]
    \centering
    \includegraphics[width=\linewidth]{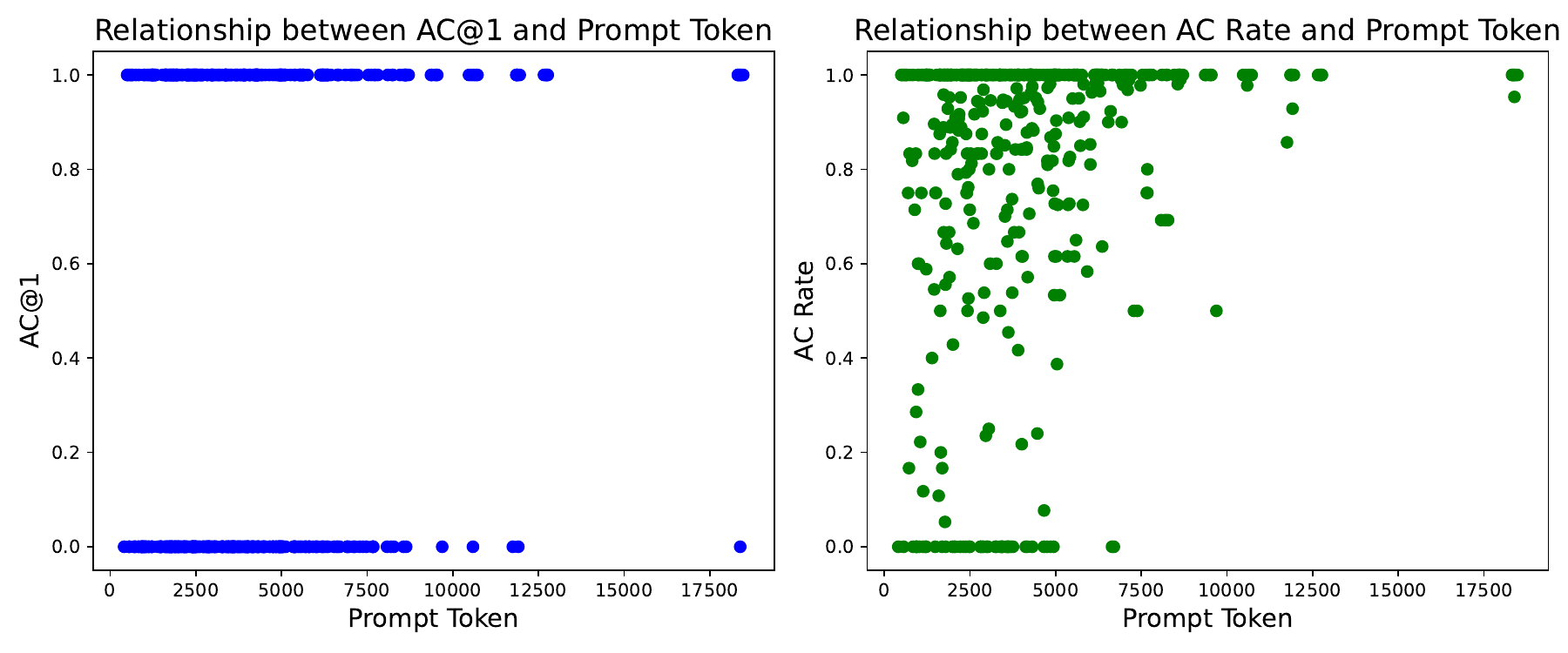}
    \caption{\textbf{Relationship between Prompt Length and Model Performance}. } 
    \label{fig:promptlen}
\end{figure}

\textsc{CoreCodeBench} also exhibits strong stability with respect to variations in prompt context size. 
As visualized in Figure~\ref{fig:promptlen}, both AC@1 and AC Rate exhibit highly scattered distributions across different prompt lengths, with no discernible trend.
Quantitative analysis using Kendall’s tau correlation coefficient~\citep{stepanov2015kendallcorrelationcoefficient} confirms this independence:
\begin{itemize}[leftmargin=10pt]
    \item Correlation between Context Size and AC@1: $\tau = 0.109$ (Negligible)
    \item Correlation between Context Size and AC Rate: $\tau = 0.153$ (Negligible)
\end{itemize}
These near-zero correlations indicate that the benchmark's difficulty is driven by semantic complexity rather than mere token count, ensuring fair evaluation across varying context windows.
This observation validates our design choice in atomic task generation, where we explicitly modulate task difficulty via \textit{Mask Length} rather than the total prompt length. It confirms that \textsc{CoreCodeBench} concerns reasoning depth, not just the ability to process long contexts.

\section{Implementation Details}\label{app:implementationDetails}

\subsection{Task Generation Configuration}
For the identification of core code blocks, we enforce a minimum threshold of 10 lines of code to filter out trivial snippets and ensure sufficient reasoning complexity. 
For composite task generation, we configure the subgraph sampling parameters as follows:
\begin{itemize}[leftmargin=10pt]
    \item \textbf{Standard Setting}: $\nu=6$ (task quantity), $d=3$ (dependency depth).
    \item \textbf{Difficult Setting}: $\nu=\infty$ (unbounded quantity), $d=3$.
\end{itemize}
The depth constraint ($d=3$) is selected to balance complexity with the typical call-stack depth observed in real-world engineering.

\subsection{Model Inference}
To ensure reproducibility, we employ greedy decoding (temperature $= 0$, top\_p $= 1.0$) for all models. 
For open-weights models, we utilize the \texttt{vLLM} library~\citep{kwon2023efficientmemorymanagementlarge} for efficient inference.
Code blocks are extracted from model responses using standard Markdown regex patterns. 
To mitigate formatting artifacts, we apply heuristic post-processing, including indentation repair and function header alignment, maximizing the executability of the generated code.

\subsection{Execution Environment}
All evaluations are conducted within pre-configured Docker containers to guarantee isolation and reproducibility. 
These environments encapsulate all repository-specific dependencies and are designed to run on standard CPU instances (verified on 4 vCPU, 16GB RAM). 
To prevent resource exhaustion, we enforce a strict 120-second timeout for each test execution. Dockerfiles and setup scripts will be open-sourced upon publication.

\section{Human Annotation} \label{app:humanAnno}
To validate the quality of \textsc{CoreCodeBench}, we conduct a rigorous manual inspection on a sampled subset of tasks.

\subsection{Human Annotator}
We recruit three professional annotators, all holding Bachelor's degrees or higher in Computer Science or Software Engineering, with a minimum of two years of Python development experience. 
Annotators are compensated at competitive market rates in compliance with local labor regulations. To prevent fatigue-induced errors, the daily workload is strictly capped at 8 hours.

\subsection{Annotation Workflow}
The annotation process follows a strict Review-Verify-Decision protocol. We present detailed manual annotation criteria in Table~\ref{lst:criteria}.

\subsection{Quality Control}
To ensure reliability, we implement a rigorous quality control mechanism. 
Specifically, 50\% of the annotated data undergoes cross-verification by a second annotator or a quality inspector from the author team. 
Disagreements are resolved through consensus meetings. 
This process yields an Inter-Annotator Agreement (IAA) rate exceeding 95\%, demonstrating high consistency in our validation standards.

\section{Fine-tuning Validation Details} \label{app:ft_exp}

To empirically validate that CoreCodeBench provides high-quality, canonical supervision, we conduct a controlled fine-tuning experiment. This section details the data distribution, experimental setup, and analysis of the results.

\subsection{Data Statistics and Split}
We evaluate the model using a repository-level split to ensure no cross-project data leakage. Out of the 12 repositories in CoreCodeBench, we select \texttt{skfolio} as the held-out test set due to its comprehensive coverage of all three atomic task types (BugFix, Development, and TDD), while the remaining 11 repositories are used for training. Table~\ref{tab:data_dist} provides a summary of the task distribution across the entire benchmark used in this experiment.

\begin{table}[h]
\centering
\caption{Task distribution for the fine-tuning experiment.}
\label{tab:data_dist}
\scriptsize
\begin{tabular}{lcccc}
\toprule
\textbf{Split} & \textbf{BugFix} & \textbf{Development} & \textbf{TDD} & \textbf{Total} \\ \midrule
Training & 297 & 464 & 222 & 983 (89.04\%) \\
Test  & 18 & 47 & 56 & 121 (10.96\%) \\ \midrule
\textbf{Total} & \textbf{315} & \textbf{511} & \textbf{278} & \textbf{1104} \\ \bottomrule
\end{tabular}
\end{table}

\subsection{Training Methodology and Hyperparameters}
The Supervised Fine-Tuning (SFT) of Qwen3-8B is conducted using a full-parameter methodology with the DeepSpeed ZeRO-3 framework. The configuration ensures the model can handle long-context repository-level information:
\begin{itemize}[leftmargin=10pt]
    \item Optimization: AdamW optimizer with a peak learning rate of $1.0 \times 10^{-5}$ and a cosine decay scheduler.
    \item Hardware \& Precision: Training is performed in BFloat16 precision. We use a per-device batch size of 1 with a gradient accumulation factor of 4, resulting in an effective global batch size of 32 across 8 GPUs.
    \item Sequence Length: The maximum sequence length is set to 32,768 tokens to accommodate complex multi-file contexts.
\end{itemize}

\subsection{Fine-tuning Results and Analysis}
Table~\ref{tab:ft_result} compares the performance of the base model and the fine-tuned version on the test set. The results demonstrate a significant performance leap. In the Development task, the model improves from 0.00\% to 19.15\% in AC@1, demonstrating that the model successfully acquired generalizable repository-level reasoning capabilities to navigate the unseen \texttt{skfolio} architecture. Notably, the BugFix AC Rate nearly doubles (7.78\% $\rightarrow$ 15.58\%), indicating that the model's generated patches become significantly more aligned with the ground truth and project constraints. These improvements on a completely unseen repository confirm that CoreCodeBench provides a high-quality, canonical supervision signal that fosters genuine generalization in software engineering tasks.

\begin{table}[h]
\caption{\textbf{Model Performance before and after SFT.} We report AC@1 and AC Rate (\%) for Development, Bug-Fix, and TDD.}
\label{tab:ft_result}
\centering
\scriptsize
\setlength{\tabcolsep}{4pt} 
\renewcommand{\arraystretch}{1.2}
\resizebox{\linewidth}{!}{
\begin{tabular}{lcccccc}
\toprule
 \multirow{2}{*}{\textbf{Model}} & \multicolumn{2}{c}{\textbf{Development}} & \multicolumn{2}{c}{\textbf{BugFix}} & \multicolumn{2}{c}{\textbf{TDD}} \\ 
\cmidrule(lr){2-3} \cmidrule(lr){4-5} \cmidrule(lr){6-7} 
& \textbf{AC@1} & \textbf{Rate} & \textbf{AC@1} & \textbf{Rate} & \textbf{AC@1} & \textbf{Rate} \\ 
\midrule
Qwen3-8B  & 0.00 & 49.93 & 0.00 & 7.78 & 16.07 & 68.00 \\
Qwen3-8B-FineTuned & \textbf{19.15} & \textbf{59.02} & 0.00 & \textbf{15.58} & \textbf{21.43} & 64.51 \\
\bottomrule
\end{tabular}}
\end{table}

\section{Confidence Intervals and Results for Additional LLMs}\label{app:evaluation-results}
Tables~\ref{tab:SingleFunction-detail} and \ref{tab:MultiFunction-detail} provide the detailed evaluation results and confidence intervals across various categories and problem types. Notably, we extend our reporting to include 12 models in these appendices to increase the number of sample points for the analysis in Section~\ref{subsec:rq4_validity}, thereby enhancing the statistical reliability of the ranking consistency results. The methodology used to compute these intervals is described below.

\subsection{Confidence Interval Computation}
\label{app:repo_equal_ci}

For a given metric (e.g., \emph{Rate} or \emph{AC@1}) and task category, we obtain per-instance scores $\{x_{r,i}\}_{i=1}^{n_r}$ for each repository $r\in\{1,\dots,R\}$. To avoid over-weighting repositories with more instances, we adopt repo-equal weighting.

\paragraph{Repo-equal mean.}
We first compute the per-repository mean
\[
\bar{x}_r=\frac{1}{n_r}\sum_{i=1}^{n_r} x_{r,i},
\]
and then define the overall mean as the unweighted average over repositories
\[
\mu=\frac{1}{R}\sum_{r=1}^{R}\bar{x}_r.
\]

\paragraph{Basic (CLT) 95\% CI under repo-equal weighting.}
For each repository $r$, we estimate the standard error (SE) of its mean by
\[
\mathrm{SE}_r=\frac{s_r}{\sqrt{n_r}},\qquad
s_r=\sqrt{\frac{1}{n_r-1}\sum_{i=1}^{n_r}\left(x_{r,i}-\bar{x}_r\right)^2}.
\]
We then combine repository-level uncertainties (treating repositories as independent) as
\[
\mathrm{SE}=\sqrt{\frac{1}{R^2}\sum_{r=1}^{R}\mathrm{SE}_r^2},
\]
and report the normal-approximation 95\% confidence interval
\[
\mu \pm z_{0.975}\,\mathrm{SE},
\]
where $z_{0.975}\approx 1.96$.

\begin{table*}[t] 
\caption{Leaderboard on Atomic Tasks. We report AC@1 and Rate (\%) with their 95\% CI for Overall, Development, BugFix, and TDD, where CI denotes the 95\% confidence interval half-width (margin) shown after the ``$\pm$''.}
\label{tab:SingleFunction-detail}

\centering
\scriptsize
\setlength{\tabcolsep}{1.8pt} 
\renewcommand{\arraystretch}{1.2}
\resizebox{\linewidth}{!}{%
\begin{tabular}{clcccccccc}
\toprule
\multirow{2}{*}{\textbf{Type}} & \multirow{2}{*}{\textbf{Model}} 
& \multicolumn{2}{c}{\textbf{Overall}} 
& \multicolumn{2}{c}{\textbf{Development}} 
& \multicolumn{2}{c}{\textbf{BugFix}} 
& \multicolumn{2}{c}{\textbf{TDD}} \\
\cmidrule(lr){3-4} \cmidrule(lr){5-6} \cmidrule(lr){7-8} \cmidrule(lr){9-10}
& & \textbf{AC@1 $\pm$ CI} & \textbf{Rate $\pm$ CI} 
  & \textbf{AC@1 $\pm$ CI} & \textbf{Rate $\pm$ CI} 
  & \textbf{AC@1 $\pm$ CI} & \textbf{Rate $\pm$ CI} 
  & \textbf{AC@1 $\pm$ CI} & \textbf{Rate $\pm$ CI} \\
\midrule

\multirow{6}{*}{\rotatebox{90}{API}}
& Gemini-3-Pro~\citep{gemini3} 
  & 68.03 $\pm$ 2.75 & 85.36 $\pm$ 1.83 
  & 67.24 $\pm$ 4.92 & 87.72 $\pm$ 2.60 
  & 62.77 $\pm$ 7.61 & 78.58 $\pm$ 5.46 
  & 73.39 $\pm$ 6.28 & 89.33 $\pm$ 3.68 \\

& Claude-4.5-Opus~\citep{claude45opus} 
  & 61.87 $\pm$ 2.87 & 78.54 $\pm$ 2.22 
  & 68.26 $\pm$ 4.73 & 86.56 $\pm$ 2.66 
  & 51.66 $\pm$ 7.79 & 63.25 $\pm$ 6.81 
  & 61.24 $\pm$ 6.69 & 83.92 $\pm$ 3.87 \\

& GPT-5.2~\citep{gpt52} 
  & 58.70 $\pm$ 2.91 & 78.70 $\pm$ 2.12 
  & 57.66 $\pm$ 5.16 & 81.80 $\pm$ 3.11 
  & 53.56 $\pm$ 7.70 & 70.92 $\pm$ 6.41 
  & 67.35 $\pm$ 6.60 & 83.27 $\pm$ 4.09 \\

& Claude-3.7-Sonnet~\citep{anthropic2025claude37} 
  & 43.75 $\pm$ 2.93 & 70.01 $\pm$ 2.35 
  & 52.81 $\pm$ 5.20 & 82.94 $\pm$ 3.18 
  & 27.90 $\pm$ 7.45 & 45.63 $\pm$ 7.38 
  & 45.11 $\pm$ 12.74 & 70.30 $\pm$ 11.70 \\

& Seed-1.6~\citep{seed16} 
  & 39.22 $\pm$ 2.88 & 64.69 $\pm$ 2.50 
  & 39.70 $\pm$ 5.05 & 69.35 $\pm$ 4.09 
  & 41.57 $\pm$ 8.02 & 64.70 $\pm$ 6.73 
  & 37.93 $\pm$ 6.51 & 52.90 $\pm$ 5.78 \\

& Qwen3-Max-Instruct~\citep{qwen2025qwen3} 
  & 47.74 $\pm$ 2.95 & 70.95 $\pm$ 2.39 
  & 58.72 $\pm$ 5.14 & 84.30 $\pm$ 3.07 
  & 29.54 $\pm$ 7.70 & 41.15 $\pm$ 7.59 
  & 47.18 $\pm$ 12.86 & 75.80 $\pm$ 11.64 \\

\midrule

\multirow{6}{*}{\rotatebox{90}{Open-Source}}
& Kimi-K2~\citep{kimik2} 
  & 47.83 $\pm$ 2.95 & 73.47 $\pm$ 2.27 
  & 58.95 $\pm$ 5.04 & 86.34 $\pm$ 2.67 
  & 29.57 $\pm$ 5.86 & 45.63 $\pm$ 6.49 
  & 50.99 $\pm$ 6.77 & 79.45 $\pm$ 4.00 \\

& Deepseek-V3.2~\citep{deepseekv32} 
  & 53.89 $\pm$ 2.94 & 79.08 $\pm$ 2.03 
  & 62.06 $\pm$ 5.09 & 87.16 $\pm$ 2.58 
  & 41.00 $\pm$ 7.60 & 60.83 $\pm$ 6.99 
  & 57.59 $\pm$ 6.70 & 83.04 $\pm$ 4.01 \\

& Qwen3-Coder-480B-A3B-Instruct~\citep{qwencoder} 
  & 50.45 $\pm$ 2.95 & 76.77 $\pm$ 2.12 
  & 57.35 $\pm$ 5.21 & 83.96 $\pm$ 3.20 
  & 36.46 $\pm$ 7.66 & 55.03 $\pm$ 7.05 
  & 54.19 $\pm$ 6.80 & 83.96 $\pm$ 3.57 \\

& GLM-4.6~\citep{glm46} 
  & 51.63 $\pm$ 2.95 & 78.94 $\pm$ 2.01 
  & 52.56 $\pm$ 5.24 & 83.26 $\pm$ 3.18 
  & 42.46 $\pm$ 7.84 & 63.68 $\pm$ 6.83 
  & 65.23 $\pm$ 6.49 & 87.33 $\pm$ 3.20 \\

& Minimax-M2~\citep{minimaxm2} 
  & 22.01 $\pm$ 2.45 & 46.40 $\pm$ 2.64 
  & 23.12 $\pm$ 4.35 & 53.50 $\pm$ 4.60 
  & 18.56 $\pm$ 5.66 & 30.84 $\pm$ 6.63 
  & 33.17 $\pm$ 5.67 & 61.96 $\pm$ 5.17 \\

& Llama-3.1~\citep{meta2024introducingllama31} 
  & 24.64 $\pm$ 2.54 & 58.09 $\pm$ 2.46 
  & 33.40 $\pm$ 5.20 & 69.15 $\pm$ 3.82 
  & 17.37 $\pm$ 6.37 & 33.19 $\pm$ 6.79 
  & 32.25 $\pm$ 12.69 & 69.51 $\pm$ 11.57 \\

\bottomrule
\end{tabular}
}
\end{table*}

\begin{table*}[t] 

\caption{Leaderboard on Composite Tasks. We report AC@1 and Rate (\%) with their 95\% CI, where CI denotes the 95\% confidence interval half-width (margin) shown after the ``$\pm$''.}

\label{tab:MultiFunction-detail}

\centering

\scriptsize

\setlength{\tabcolsep}{1.8pt} 
\renewcommand{\arraystretch}{1.2}

\resizebox{\linewidth}{!}{%
\begin{tabular}{clcccccccc}
\toprule
\multirow{2}{*}{\textbf{Type}} & \multirow{2}{*}{\textbf{Model}} & \multicolumn{2}{c}{\textbf{Overall}} & \multicolumn{2}{c}{\textbf{Development}} & \multicolumn{2}{c}{\textbf{BugFix}} & \multicolumn{2}{c}{\textbf{TDD}} \\
\cmidrule(lr){3-4} \cmidrule(lr){5-6} \cmidrule(lr){7-8} \cmidrule(lr){9-10}
& & \textbf{AC@1 $\pm$ CI} & \textbf{Rate $\pm$ CI} & \textbf{AC@1 $\pm$ CI} & \textbf{Rate $\pm$ CI} & \textbf{AC@1 $\pm$ CI} & \textbf{Rate $\pm$ CI} & \textbf{AC@1 $\pm$ CI} & \textbf{Rate $\pm$ CI} \\
\midrule

\multirow{6}{*}{\rotatebox{90}{API}}
& Gemini-3-Pro~\citep{gemini3} & 20.73 $\pm$ 4.39 & 40.30 $\pm$ 4.55 & 18.45 $\pm$ 5.91 & 35.08 $\pm$ 6.69 & 0.00 $\pm$ 0.00 & 15.99 $\pm$ 16.62 & 20.67 $\pm$ 7.31 & 42.16 $\pm$ 8.03 \\
& Claude-4.5-Opus~\citep{claude45opus} & 16.46 $\pm$ 4.02 & 26.83 $\pm$ 4.21 & 16.06 $\pm$ 6.90 & 27.09 $\pm$ 6.96 & 0.00 $\pm$ 0.00 & 13.85 $\pm$ 15.08 & 17.80 $\pm$ 5.10 & 25.51 $\pm$ 5.26 \\
& GPT-5.2~\citep{gpt52} & 9.15 $\pm$ 3.12 & 16.15 $\pm$ 3.56 & 3.82 $\pm$ 3.53 & 9.34 $\pm$ 4.45 & 0.00 $\pm$ 0.00 & 13.85 $\pm$ 15.08 & 13.76 $\pm$ 6.19 & 20.34 $\pm$ 6.52 \\
& Claude-3.7-Sonnet~\citep{anthropic2025claude37} & 12.80 $\pm$ 3.62 & 29.83 $\pm$ 4.09 & 13.21 $\pm$ 5.92 & 29.23 $\pm$ 6.62 & 0.00 $\pm$ 0.00 & 16.99 $\pm$ 16.48 & 12.16 $\pm$ 5.72 & 28.93 $\pm$ 7.14 \\
& Seed-1.6~\citep{seed16} & 4.88 $\pm$ 2.33 & 15.30 $\pm$ 3.24 & 0.30 $\pm$ 0.58 & 7.34 $\pm$ 4.05 & 0.00 $\pm$ 0.00 & 13.85 $\pm$ 15.08 & 7.61 $\pm$ 4.59 & 20.87 $\pm$ 6.70 \\
& Qwen3-Max-Instruct~\citep{qwen2025qwen3} & 10.37 $\pm$ 3.30 & 28.92 $\pm$ 3.97 & 6.01 $\pm$ 4.31 & 23.93 $\pm$ 5.81 & 0.00 $\pm$ 0.00 & 15.99 $\pm$ 16.62 & 13.83 $\pm$ 6.25 & 31.75 $\pm$ 7.02 \\

\midrule

\multirow{6}{*}{\rotatebox{90}{Open-Source}}
& Kimi-K2~\citep{kimik2} & 8.84 $\pm$ 3.08 & 22.88 $\pm$ 3.76 & 8.78 $\pm$ 4.61 & 25.06 $\pm$ 6.61 & 0.00 $\pm$ 0.00 & 2.14 $\pm$ 7.00 & 5.63 $\pm$ 3.75 & 16.21 $\pm$ 5.21 \\
& Deepseek-V3.2~\citep{deepseekv32} & 14.02 $\pm$ 3.76 & 31.68 $\pm$ 4.11 & 15.36 $\pm$ 5.93 & 32.48 $\pm$ 7.31 & 0.00 $\pm$ 0.00 & 13.85 $\pm$ 15.08 & 9.98 $\pm$ 4.90 & 27.80 $\pm$ 6.21 \\
& Qwen3-Coder-480B-A3B-Instruct~\citep{qwencoder} & 9.45 $\pm$ 3.17 & 21.30 $\pm$ 3.68 & 12.80 $\pm$ 6.00 & 25.89 $\pm$ 6.26 & 0.00 $\pm$ 0.00 & 18.66 $\pm$ 15.74 & 8.77 $\pm$ 4.35 & 21.01 $\pm$ 6.50 \\
& GLM-4.6~\citep{glm46} & 11.89 $\pm$ 3.51 & 27.45 $\pm$ 4.02 & 6.76 $\pm$ 4.13 & 24.77 $\pm$ 5.82 & 0.00 $\pm$ 0.00 & 15.99 $\pm$ 16.62 & 15.91 $\pm$ 6.56 & 31.32 $\pm$ 6.94 \\
& Minimax-M2~\citep{minimaxm2} & 2.13 $\pm$ 1.57 & 8.76 $\pm$ 2.54 & 0.77 $\pm$ 1.51 & 5.30 $\pm$ 2.94 & 0.00 $\pm$ 0.00 & 0.00 $\pm$ 0.00 & 2.49 $\pm$ 2.17 & 10.99 $\pm$ 4.33 \\
& Llama-3.1~\citep{meta2024introducingllama31} & 6.10 $\pm$ 2.59 & 18.98 $\pm$ 3.34 & 5.28 $\pm$ 3.67 & 18.67 $\pm$ 4.79 & 0.00 $\pm$ 0.00 & 17.66 $\pm$ 15.88 & 6.24 $\pm$ 4.19 & 16.21 $\pm$ 4.95 \\

\bottomrule
\end{tabular}
}
\end{table*}

\section{Full Prompts for \textsc{CorePipe}}\label{app:repopreprocess}
\subsection{Prompts of Repository Preprocessing}
We employ Claude3.5 to analyze the file structure of each repository and automatically identify the main test directories and the source code directories. The prompt used is shown below:

\begin{lstlisting}[breaklines=true]
Below is the file tree of a code repository:
{file_structure}

Please analyze the given file names and paths to identify the corresponding relationships between source code and test files (paying special attention to paths containing /test/, /unit/, or /unittest/), and provide the output in JSON format. Note that the correspondence must be based on root path relationships (for example, if both transformers/test/repo/ and transformers/test/utils/ exist, select transformers/test/). If specific unit tests exist, the relationship should be detailed to the unit test folder (such as unit), and the correspondence can tolerate some missing files as long as the files generally correspond. If there are no similar corresponding relationships, please output an empty JSON object.

Example Input:
```
- mlflow/gateway.py
- mlflow/gateway/providers.py
- mlflow/gateway/schemas.py
- mlflow/gemini.py
- mlflow/groq.py
- tests/test_gateway.py
- tests/gateway/test_providers.py
- tests/gateway/test_schemas.py
- mlflow/core/pipeline.py
- mlflow/core/pipeline/graph.py
- core_tests/pipeline.py
- core_tests/pipeline/graph.py
```
Example Output:
```
{
    "repo_name": "mlflow",
    "testcase_dir_mapping":{
        "mlflow/": "tests/",
        "mlflow/core/": "core_tests/",
    },
}
```

Note that after obtaining the mapping, perform a check to merge paths for repeated occurrences of upper-level directories; remove paths for non-core code segments (such as cli, community, _sdk, _cli/, etc.); and merge paths in cases where possible. For example:
```
{
    "repo_name": "langchain",
    "testcase_dir_mapping": {
        "libs/cli/langchain_cli/": "libs/cli/tests/unit_tests/",
        "libs/community/langchain_community/": "libs/community/tests/unit_tests/",
        "libs/core/langchain_core/": "libs/core/tests/unit_tests/",
        "libs/langchain/langchain/": "libs/langchain/tests/unit_tests/",
        "libs/partners/anthropic/langchain_anthropic/": "libs/partners/anthropic/tests/unit_tests/",
        "libs/partners/chroma/langchain_chroma/": "libs/partners/chroma/tests/unit_tests/",
        "libs/partners/exa/langchain_exa/": "libs/partners/exa/tests/unit_tests/",
        "src/transformers/": "tests/",
        "src/transformers/models/": "tests/models/",
        "src/transformers/benchmark/": "tests/benchmark/",
        "inference_sdk/": "tests/inference_sdk/unit_tests/",

        "inference/core/": "tests/inference/unit_tests/core/",
        "inference/enterprise/": "tests/inference/unit_tests/enterprise/", 
        "inference/models/": "tests/inference/unit_tests/models/",
        "inference/core/workflows/": "tests/workflows/unit_tests/"
    }
}
```
No explanations are needed, just output in JSON format and using ``` ```.
```
{
    "repo_name": "langchain",
    "testcase_dir_mapping": {
        "libs/core/langchain_core/": "libs/core/tests/unit_tests/",
        "libs/langchain/langchain/": "libs/langchain/tests/unit_tests/",
        "src/transformers/": "tests/",
        "inference/": "tests/inference/unit_tests/"
    }
}
```
\end{lstlisting}
Our approach supports cases with multiple root directories, such as repositories such as \texttt{langchain}, which contain both source code and embedded packages (e.g., \texttt{langchain} and \texttt{langchain\_core}).

After determining the main test and source directories, we traverse all files within these directories to establish fine-grained mappings between individual test files and their corresponding source files. Once valid mappings are identified, we execute the test files in the environment to verify their usability. Additionally, we record the number of test cases in each test file, which is later used to calculate the AC Rate.

We also use Claude-3.7-Sonnet to choose core code for the target function, requiring it to contain key functionality, external calls, algorithms, or core logic. Code consisting only of simple assignments or mechanical processing is excluded. Prompts for core code selection is shown below:

\begin{lstlisting}[breaklines=true]
The definition of key code blocks is as follows:  
- Code sections that implement the main functionality of the function and directly determine whether the function can achieve its intended goal;
- Code sections whose execution efficiency significantly impacts the function's performance.

Based on the code of function {func}, identify the key code blocks within block {recur}, and output the block_ids of its sub-key code blocks. The total number of lines in the selected code blocks should not exceed 60 lines, so please select carefully to ensure the most important parts are chosen.

Output format:  
If you select multiple **consecutive** blocks, please output a list of block_ids:
```python
blocks = ["blockid1", "blockid2", ...]
```
If the function is relatively simple and only contains initialization or return statements, it means there are no key code blocks. In this case, please output:
```python
blocks = None
```
Do not include additional comments in the code section; only output the blockid(s).

Please select key code blocks from the sub-blocks of the {recur} code block.

Function code:  
{code['func_code']}

Function block information:  
{code['block_info']}
\end{lstlisting}

To validate the capability of LLMs to select core code, we randomly sample 50 generated problems for manual inspection and find that all samples (100\%) meets our standards for core code selection, demonstrating the accuracy and dependability of our process.

\subsection{Prompts for Single-Dev. Generation}\label{app:dev-prompt}

\textbf{Prompt for Explanation Generation}

\begin{lstlisting}[breaklines=true]
Please analyze the provided code block based on its context, and output its functionality using concise language in the given format (do not include extra content):
1. **Purpose**
   Describe the main goal of the code block and its role within the entire program. Specifically, what is its responsibility within the current function?
2. **Logic**
   Elaborate on the core logic and operational process of the code block. For all conditional branches (if statements), explain them one by one.
   If complex variable updates are involved, use Markdown format for formulas to represent these mathematical calculations.
   If variables from previous sections of the code block are used, try to describe using their variable names, enclosing them in backticks. Functions should be enclosed in backticks as well, and can be in the form `function_name(arguments)` or `function_name`, without causing ambiguity such as `function_name()` which might lead to misunderstanding.
3. **Exceptions**
   If the code block under analysis throws exceptions, explain its exceptional cases and types. If no exceptions are thrown within the code block, state "None."
4. **Variable Assignments**
   Given the variable list, provide the specific significance and role of the computed variable in the code block in list form.
   If any variables are incorrectly identified or unused in subsequent sections of code, these can be directly removed. 
   If the variable list is missing any modified variable (such as `self.blockid_list.append(block)`), please add it to the list.
   Variable list: {variable_list}

### Sample Output:
1. **Purpose**
   Parse the target string to extract key information. The target string is in the format `blocks = ["blockid1", "blockid2", ...]`. This code block extracts all valid blockids, generating a new list of strings.
2. **Logic**
   Uses regular expressions (re library) to extract blockid list from the target string, then iterates the list, verifies each blockid's existence in the database, and stores them converted to integer type in a new list.
3. **Exceptions**
   - `ValueError`: If the target string has an incorrect format, making it unable to extract a valid blockid list, this exception is thrown.
4. **Variable Assignments**
   - `self.blockid_list`: Stores extracted and validated blockids

### Code Block to be Analyzed:
{key_block}

### Contextual Information of Code Block:
{class_code}
\end{lstlisting}
\textbf{Prompt for Refinement}
\begin{lstlisting}[breaklines=true]
The code reviewers found the generated code explanation has the following issues:
{response}

Please modify the current code explanation based on the content of the code block and the reviewers' suggestions, and output it according to the specified format, **do not include extra content**.
### Code Block to be Analyzed:
{key_block}

### Current Code Explanation:
{explanation}

### Output Requirements:
1. **Purpose**
   Describe the main goal of the code block and its role within the entire program. Specifically, what is its responsibility within the current function?
2. **Logic**
   Elaborate on the core logic and operational process of the code block. For all conditional branches (if statements), explain them one by one. 
   If complex variable updates are involved, use Markdown format for formulas to represent these mathematical calculations.
   If variables from previous sections of the code block are used, try to describe using their variable names, enclosing them in backticks.
3. **Exceptions**
   If the analyzed code block throws exceptions (using `raise` statements, excluding `except` statements), explain its exceptional cases and types. If no exceptions are thrown within the code block, state "None."
4. **Variable Assignments**
   Using the provided variable list, describe the specific significance and role of the computed variable in the code block in list form.
   If there are any erroneously identified variables (e.g., those not used later in the code), you may directly remove these. If the variable list is missing any modified variable (such as `self.blockid_list.append(block)`), please add it to the list.

### Sample Output:
1. **Purpose**
   Parse the target string to extract key information. The target string format is `blocks = ["blockid1", "blockid2", ...]`. This code block extracts all valid blockids and generates a new list of strings.
2. **Logic**
   Uses regular expressions (re library) to extract blockid list from the target string, then iterates the list, verifies each blockid's existence in the database, and stores them converted to integer type in a new list.
3. **Exceptions**
   - `ValueError`: If the target string has an incorrect format making it impossible to extract a valid blockid list, this exception is thrown.
4. **Variable Assignments**
   - `self.blockid_list`: Stores extracted and validated blockids.
\end{lstlisting}

\begin{table*}[t] 
\begin{tcolorbox}[
    colback=gray!10!white, 
    colframe=gray!80!black, 
    title=\textbf{Annotation Criteria},
    fonttitle=\bfseries,
    arc=3pt
]

\normalsize 
The evaluation consists of three aspects: Readability, Accuracy, and Completeness. Each aspect is scored on a three-level scale:
\begin{itemize}[leftmargin=25pt, topsep=5pt]
    \item \textbf{0 points:} Unusable, with obvious flaws.
    \item \textbf{1 point:} Minor flaws.
    \item \textbf{2 points:} Perfect, no flaws.
\end{itemize}
Problems with a total score of 5 or higher across the three aspects are considered qualified.

\vspace{5pt}
\hrule
\vspace{10pt}

\paragraph*{1. Readability}
Comments should be clear, concise, and easy to understand, enabling developers to quickly grasp the code’s functionality and purpose.
\begin{itemize}[leftmargin=25pt]
    \item Comments are clear and easily readable by software engineers.
    \item Sentences are fluent, with no typographical errors.
    \item Formatting complies with Markdown standards.
    \item Comments express the code’s functionality or requirements accurately using minimal wording, avoiding verbosity while ensuring clarity.
\end{itemize}

\paragraph*{2. Accuracy}
Comments must faithfully reflect the behavior of the code, ensuring that the functionality implemented based on the comments matches the actual code behavior.
\begin{itemize}[leftmargin=25pt]
    \item Comments accurately describe the code’s functionality, and the described logic matches the original code implementation.
    \item Important functional functions that need to be used are clearly indicated.
    \item Important variables that are modified (including class member variables) are listed accurately, with clear explanations.
    \item Utility function selection is correct; incorrect selection results in 0 points.
    \item Exception handling is correctly identified.
\end{itemize}

\paragraph*{3. Completeness}
Comments should cover all critical aspects of the code, especially inputs, outputs, data structures, algorithms, and edge cases, without omitting any key content that affects correct understanding.
\begin{itemize}[leftmargin=25pt]
    \item Comments cover all key aspects of the code, without missing important context (such as inputs, outputs, data structures, algorithms, edge cases).
    \item Comments are directly related to the code’s functionality and do not contain irrelevant or redundant information.
    \item Comments do not omit any branch logic or other elements that may affect correct understanding; omissions that lead to missing functionality are considered completeness issues.
\end{itemize}
\end{tcolorbox}
\caption{\textbf{Human Annotation Criteria.} } 
\label{lst:criteria}
\end{table*}

\end{document}